\documentclass[lettersize,journal]{IEEEtran}
\usepackage{amsmath,amsfonts}
\usepackage{algorithmic}
\usepackage{algorithm}
\usepackage{array}
\usepackage[caption=false,font=normalsize,labelfont=sf,textfont=sf]{subfig}
\usepackage{textcomp}
\usepackage{stfloats}
\usepackage{url}
\usepackage{verbatim}
\usepackage{graphicx}
\usepackage{cite}
\graphicspath{{figs/}{figures/}{pictures/}{images/}{./}} 

\usepackage{url}

\usepackage{mathptmx}                  

\usepackage{soul}

\usepackage[table,svgnames]{xcolor} 

\usepackage{array} 
\usepackage{booktabs} 

\usepackage{hyperref}
\usepackage[capitalize]{cleveref}

\usepackage[column=0]{cellspace} 
\newcommand{\remove}[1]{\ignorespaces}

\renewcommand{\remove}[1]{\ignorespaces}

\usepackage{graphicx}

\begin{document}

\title{The Now and Then: Integrating Current and Historical Data in Small Multiple Time Series Visualization}

\author{Sydney K. Purdue, Enrico Bertini, and Melanie Tory
\thanks{All authors are with Northeastern University.} %
\thanks{Email: purdue.sy | e.bertini | m.tory@northeastern.edu}
\thanks{Manuscript received February 7, 2025; revised September 12, 2026.}}

\markboth{Journal of \LaTeX\ Class Files,~Vol.~14, No.~8, August~2021}%
{Purdue \MakeLowercase{\textit{et al.}}: A Sample Article Using IEEEtran.cls for IEEE Journals}


\maketitle

\begin{abstract}
  Small multiple time series visualizations are often used for real-time data monitoring tasks in high-impact domains such as healthcare and manufacturing. Effective design is critical because users rely on these visualizations to monitor data from many entities, such as patients or machines, often while distracted. Users may need to rapidly appraise current values for each entity, monitoring for those that go outside an acceptable range, while also watching temporal trends. However, no design guidelines currently exist for visually emphasizing current values in historical time series represented by small multiples. Via an iterative design process informed by a review of related literature and theory on visual channels and emphasis, we present a design space for glanceable time series small multiple displays. We evaluate this space through two online empirical studies, testing against non-threshold and threshold rapid appraisal tasks. Our results provide insights into merging current value and historical data visualizations for rapid appraisal tasks in time series monitoring. For non-threshold tasks, we found that size encodings on the current value, spatially integrated into the line chart, may provide a good compromise, with 28\% response time improvement for tasks involving finding large current values and minimal interference with trend lookup tasks. More generally, integrated designs outperformed separated designs (in which the current value representation is spatially separated from the historical trend line). For threshold tasks, color threshold encodings significantly outperformed shaded band encodings.
  
  All supplemental materials are available at \url{https://osf.io/wzf6a}. 
\end{abstract}

\begin{IEEEkeywords}
Small multiples, Time series, Emphasis, Visual encoding design, Monitoring tasks, Quantitative evaluation
\end{IEEEkeywords}

\section{Introduction}
\textbf{This work explores the design of glanceable small multiple displays for rapid appraisal tasks in time series data monitoring.}
\textit{Small multiple} visualizations, where small graphs are faceted across one or more variables, are often used for \textit{data monitoring} tasks, such as monitoring the heart-rate of many patients from a nursing station or the performance of equipment in a factory from a control room. Such data monitoring use cases typically involve \textit{time series} data, or data that is updated, manually or automatically, at regular intervals, and can involve many different tasks. Here we focus particularly on \textit{rapid appraisal tasks}, where a user needs to quickly obtain information \textit{at a glance}. Both the \textit{trend} over time and the \textit{current value} are of interest. Users may also need to compare how the trend or the current value interact with \textit{thresholds}, e.g., to identify if a patient is experiencing a too high or too low heart rate. Strategic visual encodings or visual emphasis on current values and values exceeding thresholds may support these rapid appraisal tasks. Considerable visual encoding guidance exists for components of these challenges, including small multiples design \cite{archambault_animation_2011, hosseinpour_examining_2024, javed_graphical_2010, fuchs_evaluation_2013, brehmer_comparative_2020}, time series representation \cite{aigner_visualization_2023, javed_graphical_2010, franke_comparative_2022}, redundant encoding \cite{nothelfer_redundant_2017, chun_redundant_2017, nowell_graphical_2002}, and visual emphasis \cite{mairena_which_2022, mairena_baseline_2020}. However, there are no combined guidelines on how to best design current value encodings in small multiples time series visualizations. Designing such encodings is challenging because simple emphasis techniques on points in the time series may be insufficient for rapidly assessing the current values, while best practices for encoding the current value could interfere with trend understanding.


Data monitoring tasks are often secondary; workers may have many other draws on their attention. Therefore, reorienting to visualizations quickly is a major concern.
Monitoring tasks in general have been understudied in visualization \cite{sarikaya_what_2019}. Investigation is needed into how to design glanceable displays that enable rapid appraisal tasks related to both the \textit{current value} and the \textit{historical trend}, which may add complications through a combination of visual techniques. Prior research does not address the combination or interaction of these  methods, which may influence effectiveness when used together.

Monitoring tasks using time series small multiples are frequently found in high-stakes applications, such as health care, manufacturing, and power production, where effective visualization design is critical to preventing failures of important systems, costly damages, and losses of life.
For example, we are motivated by our ongoing collaboration with a hospital intensive care unit (ICU) \cite{barry_needs_2026}. An ICU may have monitors at various locations, including at a central nurses’ station, that compile the streaming vital signs from all patients currently in the unit. These monitors show the outputs of various vitals, though more complete information is available at patients’ bedsides. For healthcare staff throughout the ICU, these monitors provide quick information 
as to whether a patient is experiencing a health issue. For instance, a glance at these displays can enable healthcare workers
to rapidly assess whether an alarm is from a critical or noncritical cause (such as an equipment disconnection). Surprisingly, despite their critical importance for patient monitoring, bedside displays have seen few advancements in several decades and researchers have identified a clear need to better understand how to optimally present these data for effective monitoring \cite{de2015information}. Additional site visits performed by the research team to 4 manufacturing companies and 1 electrical grid monitoring facility provided further examples of data monitoring environments and control rooms. These organizations, among many others, have repeatedly told us that their state-of-the-art systems for data monitoring are inadequate. Surprisingly, human factors literature on process control and healthcare monitoring are quite siloed from literature on visualization. Current information representation practices in these scenarios rely heavily on threshold-based alarms, while alarm overload is a pervasive problem leading to stress and burnout~\cite{simpson2019false}. While our work does not directly address the alarm problem, it aims to reduce the monitoring burden faced by human operators by providing guidance on display design to make current values visible \textit{at a glance} while retaining their temporal context.

To propose design principles for these scenarios, we wanted to know how different design elements of time series visualizations affect people's ability to rapidly detect and identify both current values 
and trends over time, and to detect potential problems,
such as exceeding a critical threshold. Because we theorized that existing design guidelines for time series, small multiples, and emphasis may interfere with one another in combined applications, additional research was needed to find optimal design solutions. Visualizations that may be ideal for emphasizing the current value may affect tasks related to the historical trend or vice versa. To test this, we first created a design space\footnote{\textcolor{blue}{A design space is a conceptual space of possibilities that a designer could consider.}} for glanceable small multiple time series visualizations by exploring design variations. Next, we created exemplars to test experimentally. Using a set of rapid appraisal tasks representative of common tasks within our industry scenarios, we compared the designs in terms of human performance in two crowd-sourced studies. Experiment 1 examined non-threshold tasks and Experiment 2 examined comparisons to a threshold range. The design variations we tested in each experiment are shown in \cref{fig:DesignVariations}.

The key contributions of our work are:
\begin{itemize}
    \item A design space for integrating visual encodings of current and threshold values within time series small multiple visualizations for rapid appraisal tasks in monitoring applications;
    \item A task breakdown for rapid appraisal monitoring, beginning a task taxonomy for monitoring;
    \item Empirical results comparing several integrated and separated visual encodings of current values and threshold values within time series small multiples.
\end{itemize}

Our results showed that current value encodings and emphasis assisted with finding plots with high or low values but that these visual encodings can sometimes interfere with finding trends. For the design variations that we studied, integrated current value designs had lower overall interference than separated ones. For thresholds, color coding generally outperformed bands for helping participants identify if a value is inside or outside a threshold range, and importantly, current value encodings did not significantly interfere with these trend-based threshold tasks. Overall, our \textbf{Find} tasks were impacted by our design variations, while our \textbf{Lookup} tasks did not significantly vary from the control case, suggesting that our designs primarily impacted tasks that involve scanning across several small multiples rather than locating and identifying information within one. 

Our general recommendation for designers integrating current values and trends in a small-multiple setting is to prioritize integrated designs and size encoding, while remaining mindful of potential interference with trend estimation. If trend estimation is particularly important, integrated designs can be paired with simple emphasis methods without additional value encoding. In cases where thresholding is required, we suggest preferring color encoding to bands.

\section{Related Work}
\subsection{Small Multiples}
Small multiples are effective at conveying dense information across categories \cite{javed_graphical_2010, brehmer_comparative_2020, robertson_effectiveness_2008, archambault_animation_2011, fuchs_evaluation_2013}. Many studies have found that small multiples are faster than animation for tasks such as identifying change \cite{brehmer_comparative_2020} or node fluctuation \cite{archambault_animation_2011}. While animation may facilitate specific tasks, a combination of animation and small multiples may maintain advantages of both \cite{boyandin_qualitative_2012, brehmer_comparative_2020}. Javed et al. \cite{javed_graphical_2010} found that small multiples and overlapping line charts performed better than horizon and braided graphs. However, increasing the number of lines present in a line chart can make the plot difficult to read \cite{javed_graphical_2010}, which makes small multiples a better solution when there are many categories. 

Within small multiples, Fuchs et al. \cite{fuchs_evaluation_2013} found that linear time series glyph layouts were most accurate for peak and trend detection. 
For comparing values across small multiples, positional or length encodings were found to be superior to color. The authors also found that polar layouts were more accurate for comparing specific points in time, where participants made use of a clock metaphor to locate specific points. 
By emphasizing a specific point within a linear time series, it may be possible to overcome this effect and increase accuracy and efficiency, which we examine in this paper through the lens of the current value.

Hosseinpour et al. \cite{hosseinpour_examining_2024} 
found that increasing the number of multiples within a small multiples visualization linearly decreased accuracy across all tested tasks. The authors also examined a form of emphasis, by outlining frames of the small multiple grid. Motivating for our work, the authors suggest that additional research is needed for tasks involving thresholds and granular identification of specific values, such as the current value we study here. Overall, while small multiple visual encodings have been extensively studied for deep analytical tasks, our work extends this body of research to rapid appraisal monitoring tasks, specifically, how current values and threshold relationships can be emphasized without interfering with trend understanding.

\subsection{Emphasis}
Emphasis, or the technique that makes a portion of the visualization more prominent, is often used to highlight important values \cite{hall_formalizing_2016}. Much of our understanding of emphasis comes from previous research around pre-attentive processing 
and graphical perception \cite{ware_informationvisualization_2012}. 
Frequently, visual ``popout'' is understood as making some portion of the data more noticeable via changing its properties \cite{healey_attention_2012}. Visual embellishment has also been studied as a means for encouraging recall but was not found to help in visual search \cite{borgo_empirical_2012}.

Hall et al. \cite{hall_formalizing_2016} formalize emphasis. 
\textit{Intrinsic} effects vary the prominence between data points through the visual mapping, where \textit{extrinsic} effects apply effects over an existing visualization. Threshold visualizations are a notable example, where an intrinsic effect may color data points according to threshold values while an extrinsic effect would indicate threshold values through the addition of an extra line or bar. In our experiments, we have two main kinds of emphasis: the \textit{current value} in each chart, and \textit{data outside of a threshold}. 

Experiments on emphasis by Mairena et al. \cite{mairena_baseline_2020, mairena_which_2022} 
showed that to best emphasize a data point, the difference between the `focus' and `background' data points should be maximized, though this provides decreasing returns beyond a certain distance.
Blur was the most effective emphasis cue, followed by size and then color \cite{mairena_baseline_2020}. In a follow-up study, the authors found that an increased number of distractors negatively impacted time responses for size-based emphasis \cite{mairena_which_2022}, suggesting it may be more susceptible to visual clutter than other methods of emphasis design. The authors note a limitation in only emphasizing one point within a scatter plot at a time, suggesting more work remains to be done with emphasis across multiple data points.

Methods of emphasis that blur or otherwise obscure data points \cite{hall_formalizing_2016, mairena_baseline_2020, mairena_which_2022} are unsuitable to our use case, where both the trend and current value must be clearly legible. Some animations (e.g. flickering) may also be prohibitive in our settings, 
because they could be visually distracting.
Thus, in this paper, we focus on emphasis methods that are static rather than time variant.

Beecham et al. \cite{beecham_faceted_2016} propose a method of combining small multiples and representations of different kinds of data, such as temporal, spatial, and descriptive data. The authors examine a difference between superimposed and juxtaposed designs of three different kinds of data at once, and they propose that emphasis might be achieved not only by the level of abstraction used but also through the opacity of the representing marks. While this study does examine small multiples and emphasis, it does not address emphasis of individual values. 

Generalized knowledge about effective emphasis techniques provides a good starting point for our design problem. However, these techniques have not been studied in the context of supporting rapid appraisal tasks for time series monitoring.

\subsection{Visualizations for Data Monitoring}

Dashboards are a common visualization method for data monitoring 
\cite{wexler_big_2017, few_visual_2017}. Sarikaya et al. \cite{sarikaya_what_2019} identified multiple kinds of dashboards. 
Their `awareness dashboards' are the closest to small multiples for monitoring, where a display may update in real time but users are not altering the presented view. Additionally, benchmarks on `decision-making dashboards' commonly convey critical thresholds that may provide signposts for values exceeding an acceptable range.

Patient monitors are ubiquitous in in-patient healthcare settings, often featuring bedside monitors with patient-specific information and central monitors providing a multi-patient overview, but these have remained static in the past 20 years \cite{andrade_novel_2020, scheer_visualization_2022}.
Because healthcare providers are frequently interrupted, displays must enable rapid reorientation. Much research has been done into the `alarm fatigue' of healthcare providers \cite{varpio_helpful_2012, paul_designing_2016, ancker_effects_2017, lee_repeat_2019, mastrianni_pop-up_2022}, where too many constant alerts desensitize workers, potentially resulting in a disregard of important alerts. However, some research has shown that when an alarm sounds, providers will often check the displays, noting both the recent trend and current value, to assess its legitimacy and check in with patients \cite{varpio_helpful_2012, barry_needs_2026}. In these cases, the monitors factor into the workflow of alarm evaluation and resulting action or inaction, highlighting the need for informative, clear, and perceptually advantageous displays of information, motivating our work.

Data monitoring has also been studied extensively from a human factors perspective in control room settings for supervisory process control, where the emphasis is often on presenting and triaging alarms. Examples include manufacturing process control~\cite{han2007designing}, mill control~\cite{fagerlonn2017designing}, power systems~\cite{betancur2022visualization}, cyber-physical security~\cite{cobilean2023review}, and building automation~\cite{yang2014evaluation}.
Consistent themes in this literature include the need for situational awareness plus alarm triaging and the cognitive challenges of dealing with overwhelming quantities of information, which can lead to human error~\cite{domova2020guiding}.

Hu et al.~\cite{hu2018design} present a summary of standard visualization techniques that have been applied to alarm and event data dashboards. 
Research has investigated how to guide operator attention towards key events~\cite{domova2020guiding}. A common practice is the use of color to categorize alarms by severity level and guide attention to the highest priority alarms or sensor readings~\cite{han2007designing, betancur2022visualization, yang2014evaluation}, suggesting that reserving color for similar scenarios may be helpful in bridging a transition to new designs.

Few of these works investigate the visual encoding design for the monitored entities and their time series information, representing a gap in the literature. Closest is Noah et al.~\cite{noah2014evaluating}, who compared three visualizations of multiple monitored entities: a ``heat map'' (a small multiples display similar to our separated design), a ``surface chart'' (which integrated the current status of all items into a single view), and individual gauges. Results showed the value of the single-view surface chart for change detection tasks and suggested that such an overview might be a useful complement to a small multiples display like those studied here in our work.

\section{Design Space for Current Value Encoding in Time Series Small Multiples}
To explore the design space of time series small multiples, we first characterized the space itself in a way that articulates visualization approaches enabling a user to retrieve both the \textit{current value} of the data, i.e., the data point that was collected most recently, as well as the \textit{historical trend}, i.e., all visible previously collected data. More specifically, we explore how to emphasize current values without interfering in a user's ability to read a trend in a time series small multiples visualization. We focus entirely on non-interactive representations because of our focus on glanceable design to support rapid appraisal tasks; however, any of these representations could be made interactive to support deeper downstream analysis.

\begin{figure}
    \centering
    \includegraphics[width=\linewidth, alt={Examples from the design ideation stage of development, including overlaid numbers across line charts, polar layouts, factory metaphors, and floral metaphors}]{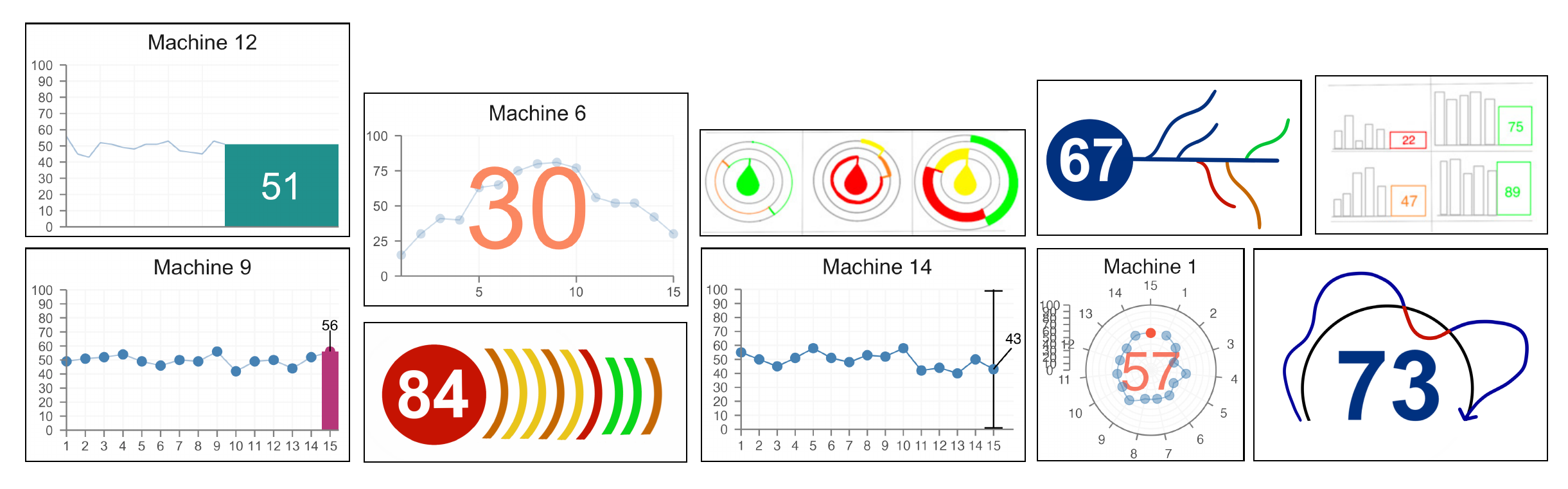}
    \caption{Examples from design ideation and development of the space. We explored designs that could include both a historical time series trend and current value. These variations helped us form the broader categories of our space by identifying common layout features across the designs.}
    \label{fig:DesignIdeation}
\end{figure}

\subsection{Development}
To develop our design space, we reviewed previous literature and iterated on existing designs from theory on visual channels. Few references exist in the visualization literature for our exact design parameters, so we surveyed time series and small multiple visualizations, emphasis design, and design for streaming data.
Application domain-specific literature provides insight into the information most relevant to users \cite{andrade_novel_2020, edrington_machine_2014}. Our ongoing field research in healthcare also informed our design process \cite{barry_needs_2026}. 

Following the collection of designs, we explored the design space through our own generated designs. Using theory on visual channels \cite{munzner_visualization_2015} and visual popout \cite{healey_attention_2012}, we systematically varied the visual encoding of the current value and historical trend. Using Munzner's visual channels \cite{munzner_visualization_2015}, we iterated through each channel, holding a channel constant for the representation of the historical trend and varying the channel used to represent the current value, thereby emphasizing the current value through its added, layered representation. For several current value/historical trend combinations, multiple methods of visualization were possible using an X-Y linear scale layout. We also examined polar layouts, but we leave a systematic review of their design to future work. Some initial explorations can be found in \cref{fig:DesignIdeation}, which shows selected examples of ideas we explored before we converged to our final design space. Each example shows a current value that is emphasized or visually encoded, either adjacent to or integrated within a time series plot.

Additionally, to capture design variations not easily surfaced through our previous method, 
we engaged in free ideation. 
One ideation question asked ``What does a visualization design look like if you assert the historical trend or current value at the detriment of the other?'' During this line of inquiry, we found it difficult to de-emphasize the historical trend in almost all cases beyond shrinking it to illegibility or transforming the line to categorical representations. The strength of linear timelines is well-established \cite{munzner_visualization_2015, aigner_visualization_2023}, and because our design requirements necessitate numerically-comparable historical data, we chose to focus on designs that varied the visual encoding and emphasis of the current value while maintaining a base legibility of the historical trend.

Similarly, we explored alternate methods of encoding trends through more metaphorical representations, where we asked ``How can a visual metaphor emphasize a current value while ensuring the trend is retrievable?'' This resulted in several alternate designs, including metaphors taken from our application areas like a machine producing bars of a bar plot along a conveyor belt (\cref{fig:DesignIdeation}, top right). We also ideated around the concept of categorical data applied on top of our variations, primarily using color as a means of establishing cutoffs or indicating when a data reading may be reaching different levels of severity. Categories could also include whether or not a value is trending in a particular direction. For this, we explored ``How can different categories of data be represented, and how can this be shown throughout the visualization?'' \cref{fig:DesignIdeation} provides selected examples.

From our gridded design generation, we had methods that varied the current value's design through one channel. However, from our additional variations, we consistently found designs that had secondary or tertiary data encodings for the representation of the current value.
Redundant encodings have shown varying effectiveness \cite{nothelfer_redundant_2017, chun_redundant_2017, nowell_graphical_2002} but have most consistently been found effective in the redundant encoding of color and shape used categorically \cite{nothelfer_redundant_2017, nowell_graphical_2002}. From our literature review, we also had variations of emphasis where data was encoded by the channel of emphasis (e.g., a point may be emphasized by making it larger than other points, but size does not need to scale with the value of the point). We also found designs that manipulated the current value's relationship to axis scales. Some resulted in an extension of the axes, where the current value was represented by a larger box that the trend flowed from. Others relocated the label of the current value, such as hovering over the center of the plot or beneath the historical trend points. This concept of ``axis-breaking'' led us our final design space.

\begin{figure*} [h]
    \centering
    \includegraphics[width=.8\paperwidth, alt={This figure shows all the design variations used in both experiments.}]{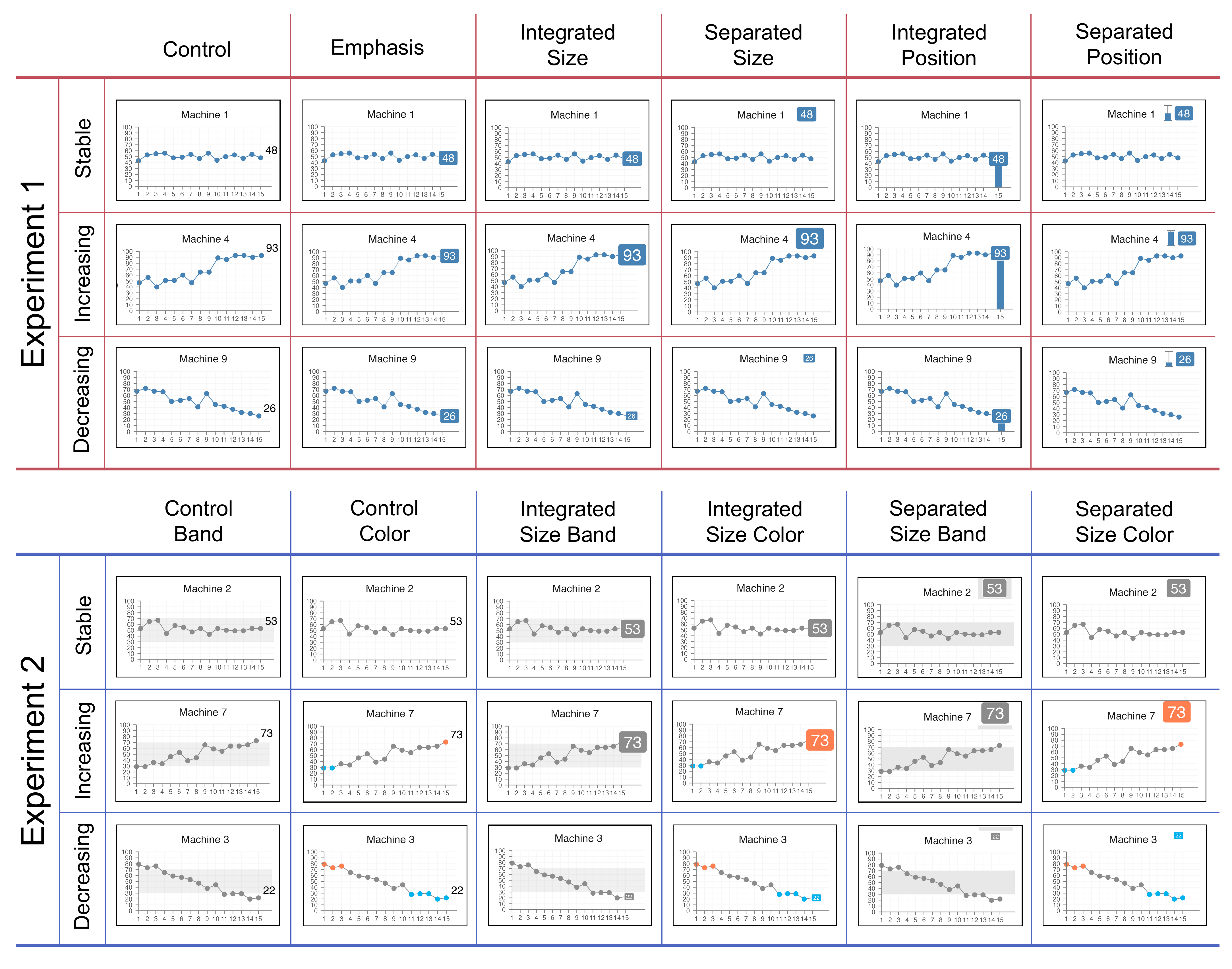}
    \caption{Design variations for both experiments across different types of trends.}
    \label{fig:DesignVariations}
\end{figure*}

\subsection{Design Space}
Our design space has four categories addressing the relationship between the representation of the current value and the historical trend present in time series small multiples. The categories are \textbf{Standard}, \textbf{Emphasis}, \textbf{Integrated}, and \textbf{Separated}. Examples of each are illustrated in Fig.~\ref{fig:DesignVariations}.

\textbf{Standard} plots do not emphasize or redundantly encode the current data value.
Current values may be labeled, but, in the case of point marks, the label is adjacent to the location and not contained within the point---hence altering the encoding of the current value point---unless all points contain their labels. Standard plots were used as controls in our experiments and are labeled as `control' conditions in Fig.~\ref{fig:DesignVariations}.

\textbf{Emphasis} plots are those where the current value is emphasized by some means that does not vary with the data value (e.g., highlighted in a different color from historical data points).
\textbf{Emphasis} plots may have labeled current values, and the design of the label may distinguish the current value from the historical trend values.

\textbf{Integrated} plots are plots where a redundant encoding is applied to the current value and the location of this encoding is embedded following the spatial coordinate system of the larger plot. The redundant channel may be size, color saturation, or another channel, but the additional encoding varies with the value of the current value. For example, if the current value is using a redundant channel of size, the representation may be larger if the current value is larger relative to the total range of the data. \textbf{Integrated} representations use the X-Y positional scales of the plot to contextualize the current value with the data of the historical trend through their spatial proximity.

\textbf{Separated} plots retain the redundant encoding properties of \textbf{integrated} plots but change the position of the current value encoding. The encoding is placed externally to the X-Y positional scales, which may be outside of the area of the plot or overlaid on the plot in a location that does not match the X-Y positioning of the current value. 

\subsection{Selection from Design Space}

Having identified these methods of special encoding of the current value to support rapid appraisal tasks in small multiple displays, we selected designs that would best explore our design space across a variety of rapid appraisal tasks.
Previous research on visual channels \cite{munzner_visualization_2015, cleveland_graphical_1984} and emphasis \cite{mairena_which_2022, mairena_baseline_2020} as well as common designs from our use cases suggested a subset of designs that could serve as exemplars, allowing us to focus our experiments on the design \textit{concepts} in the design space while minimizing replication of prior research on encoding channel effectiveness. We also considered possible visual interference as rationale for channel selection. As line charts are the most commonly used chart in our use case scenarios, we limited our experiments to these varieties. Further research is needed to establish the effect of displaying trends and current values in different ways in Cartesian space (such as bar charts, etc.).

For our first experiment, to manage the experimental scope and reduce potential participant fatigue, we needed to limit the number of current value encodings. We selected position because it is known to be the most accurate visual channel \cite{munzner_visualization_2015, quadri_survey_2022} and size because experiments \cite{mairena_baseline_2020, mairena_which_2022} have shown it to be an effective channel for visual emphasis. Color was another viable option with experimental backing, but we excluded it from our first experiment because, in monitoring displays, color is often reserved for alarm states when a threshold is exceeded.

Our second experiment was informed by our first experiment and allowed for tailoring of our design conditions in light of our previous results. When adding another layer of complexity to the tasks in the form of thresholds, we introduced color variation in two forms: a visible threshold band and a color change based on exiting a threshold area. Testing these color integrations among our other designs allowed us to test the color channel in this context while also testing our design space with these slightly modified tasks. This is also the most common use of color in monitoring system displays, accounting for greater ecological validity for our use cases.

\begin{figure}
    \centering
    \includegraphics[width=\linewidth, alt={This figure explains the design for the separated size band design variation}]{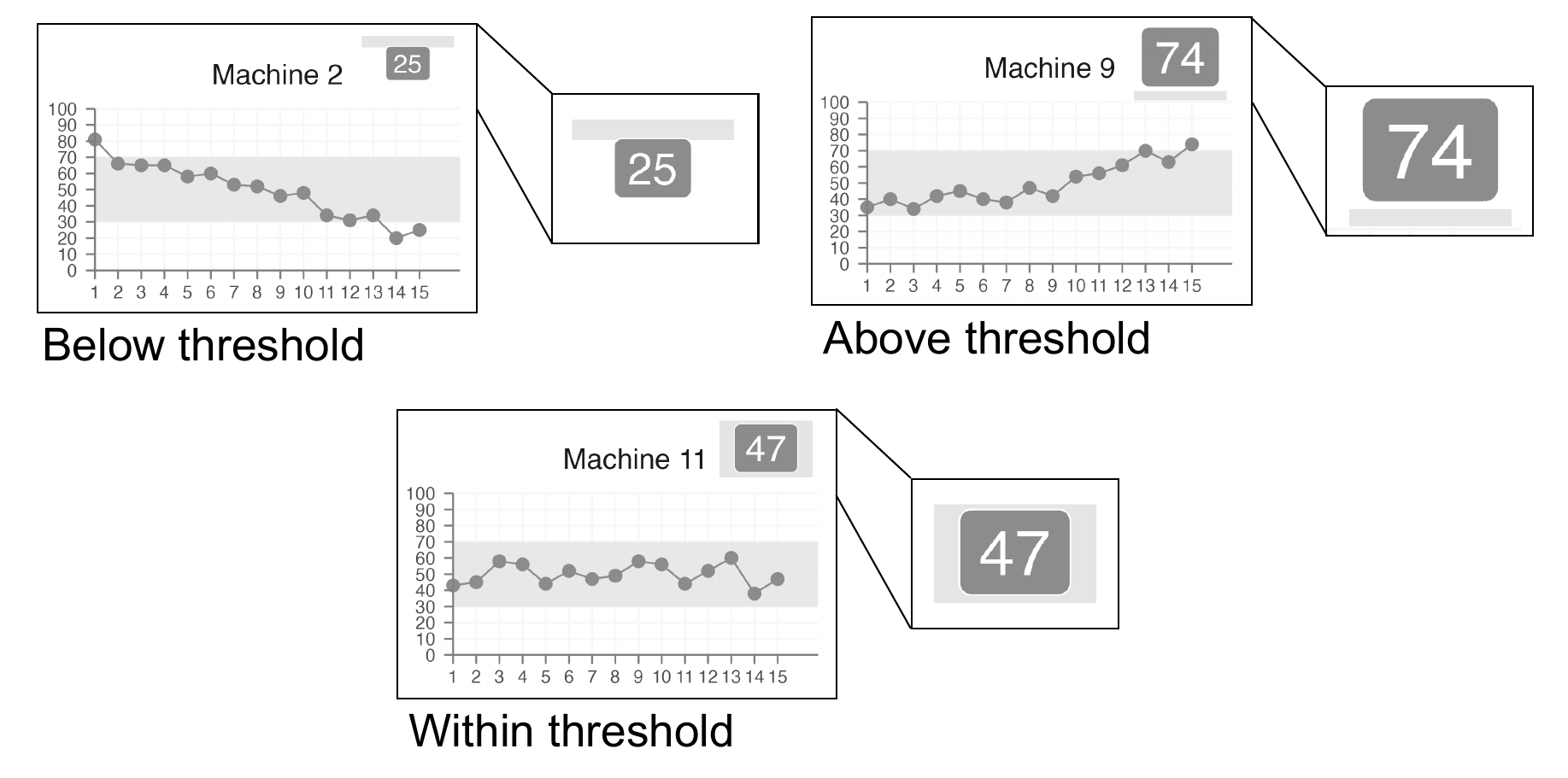 }
    \caption{Design for Separated Size condition with band threshold. The gray bar next to the separated label is placed above, underlaid, or below the label for below, within, and above threshold respectively.}
    \label{fig:SeparatedSizeBandExample}
\end{figure}

\subsection{Research Questions}
Among the variations in our design space (see Figure~\ref{fig:DesignVariations}), it was not immediately obvious which would overall perform the best for monitoring tasks. The uncertainty stemmed primarily from the possibility of perceptual \textit{interference}, which is not well understood. We have two main sources of potential interference: current value encoding and placement strategy (integrated vs. separated). For encoding, size and position are strong visual methods, suggesting that integrated designs and those with size variation or drop bars should assist with current value tasks. However, these same encodings might draw unwanted attention when the current value is not the focus, thereby interfering with trend tasks. Similarly, for placement strategy, integration may provide a more direct and unified appraisal of information than separated designs, but potentially at the cost of unwanted interference. We aim to find designs that enhance current value tasks while minimizing interference with trends.

Using our design space, we articulated the following research questions: 
(\textbf{RQ1}) Do visual encodings or visual emphasis on the current value facilitate current value-focused rapid appraisal tasks? 
(\textbf{RQ2}) Do visual encodings or visual emphasis on the current value interfere with performance on trend-focused rapid appraisal tasks?
(\textbf{RQ3}) In the context of RQ1 and RQ2, how does performance differ between separated and integrated designs?
As many rapid appraisal monitoring tasks also involve comparison of the values to a threshold range, we were interested to explore both threshold designs and tasks as well as non-threshold ones. In the next section, we describe two experiments to answer these questions.

\section{Methodology}
\label{sec:methodology}
We developed an experimental protocol with representative plots for each of our design space categories across non-threshold and threshold designs. We developed visualization tasks informed by our application areas and conducted two crowd-sourced experiments:

\textbf{Experiment 1} tested non-threshold tasks. We tested two visual channels across the integrated and separated categories in our design space. The Standard category represented a control condition and we had five test conditions, as shown in Fig.~\ref{fig:DesignVariations} (top):
\textbf{(A) Standard} (control),
\textbf{(B) Emphasis},
\textbf{(C) Integrated} size encoding,
\textbf{(D) Integrated} position encoding,
\textbf{(E) Separated} size encoding, and
\textbf{(F) Separated} position encoding.

\textbf{Experiment 2} tested tasks involving comparison to a threshold range. Based on Experiment 1's results, we dropped position encodings for the current value and focused on size. 
We tested six design variations, two controls (representing the Standard category) and four test conditions grouped by our Separated and Integrated categories. All conditions were grouped by two threshold representations: \textbf{band}, where a band is overlaid to represent the threshold range, and \textbf{color}, where data points outside the threshold range are colored differently than those inside. The conditions, shown in Fig.~\ref{fig:DesignVariations} (bottom), were:
\textbf{(A) Standard} (control) band,
\textbf{(B) Standard} (control) color,
\textbf{(C) Integrated} size encoding with band,
\textbf{(D) Integrated} size encoding with color,
\textbf{(E) Separated} size encoding with band (see Fig.~\ref{fig:SeparatedSizeBandExample}), and
\textbf{(F) Separated} size encoding with color.

\subsection{Constraints}

Our design space was developed within the framing context of streaming visualizations for rapid appraisal tasks. Within this context, update rates of streaming data can vary dramatically, from hours and minutes to seconds or sub-seconds. For simplicity of initial experiments, we choose to focus first on less frequent update rates, for which static displays can serve as an approximation for testing at-a-glance tasks. 
Many additional factors need to be considered in dynamically updating charts, such as the update rate, animated transitions or emphasis techniques, and sound-based alerts when thresholds are exceeded. By excluding these additional factors, we maintain more precise experimental control in testing our design space for this initial study. Our aim was to narrow down the space of viable visual encodings for future evaluation in a more complex, dynamic scenario.

Our experimental protocols also test exclusively on computer displays. While many monitoring tasks may take place using portable devices, we focus here on tasks that take place in aggregate across many entities being monitored simultaneously.
Our literature review and discussions with practitioners \cite{barry_needs_2026} showed that these tasks are most commonly conducted on larger displays. 
Design constraints for these scenarios are different than those on mobile devices, which may need more emphasis or larger text to accommodate the smaller screens. 

\subsection{Tasks}
As we are unaware of a visualization task taxonomy for data monitoring from previous literature, we adapted tasks from existing sets \cite{hosseinpour_examining_2024, brehmer_multi-level_2013}. We focused on rapid appraisal data monitoring activities in our target application areas: healthcare, manufacturing, and energy production. Based on our prior interviews and discussions with practitioners in these domains (e.g., \cite{barry_needs_2026}), we listed and categorized rapid appraisal monitoring tasks. 
These tasks also align with published literature \cite{brehmer_multi-level_2013, yang2014evaluation, mahmoud2003real, hosseinpour_examining_2024, quadri_survey_2022, lam2017bridging, archambault_animation_2011}. Tasks from Hosseinpour et. al \cite{hosseinpour_examining_2024} are particularly comparable to our task set.
For \textbf{Find} tasks, users need to identify properties of one or more plots from within the larger set.
\textbf{Lookup} tasks occur when users need to look up or estimate information from known individual plots. In healthcare, for example, it is common to check the display for one specific patient, which may require scanning an overview to find the patient.
Within the categories, we grouped tasks related to the current value of the streaming data vs. the historical trend of the visible data, the amount of change over the time period, and the speed of change. 

Both current value and historical trend tasks had variations that differed based on whether or not a value was outside a typical range or above/below a critical threshold. For example, in manufacturing, a machine may be producing above or below an acceptable rate; low production is a sign of malfunction, and high production may indicate reduced quality. 

Using this framework, we simplified our task list, selecting tasks that had commonalities between threshold and non-threshold variants. We focus on tasks about finding and looking up the current value and the historical trend.
Participants were asked to imagine themselves as an employee of a widget factory monitoring machines' outputs of parts per minute. Our non-threshold tasks can be seen in \cref{tab:tasks}.

\begin{table}[]
\caption{Rapid Appraisal Experimental Tasks.}
\resizebox{\columnwidth}{!}{%
\begin{tabular}[t]{@{}lll@{}}
\toprule
\multicolumn{1}{r}{} & \textbf{Find} & \textbf{Lookup} \\ \midrule
\multicolumn{3}{l}{\textbf{Non-Threshold Tasks (Experiment 1)}} \\ \midrule
\textit{\textbf{Current Value}} & \begin{tabular}[t]{@{}l@{}}\textbf{FindCurrValHigh:} Which machine has \\ the \textbf{highest} output right now?\\ \textbf{FindCurrValLow:} Which machine has \\ the \textbf{lowest} output right now?\end{tabular} & \begin{tabular}[t] {@{}l@{}} \textbf{LookupCurrVal:} What is the output of\\ Machine X now?\end{tabular} \\ \midrule
\textit{\textbf{Trend}} & \begin{tabular}[t]{@{}l@{}}\textbf{FindIncrTrend:} Find 3 plots that have \\ an overall output \textbf{increasing} over time.\\ \textbf{FindDecrTrend:} Find 3 plots that have \\ an overall output \textbf{decreasing} over time.\end{tabular} & \begin{tabular}[t]{@{}l@{}}\textbf{LookupTrend:} Is the overall output of\\ Machine X  increasing, decreasing, or\\ staying the same?\end{tabular} \\ \midrule \midrule
\multicolumn{3}{l}{\textbf{Threshold Tasks (Experiment 2)}} \\ \midrule
\textit{\textbf{Current Value}} & \begin{tabular}[t]{@{}l@{}}\textbf{FindCurrValOutsideThres:} Which \\ machines are outside of the acceptable \\ range now?\end{tabular} & \begin{tabular}[t]{@{}l@{}}\textbf{LookupCurrValThres:} Is the current\\ output of Machine X above, within,\\ or below the acceptable range of 30\\ to 70 widgets per minute?\end{tabular} \\ \midrule
\textit{\textbf{Trend}} & \begin{tabular}[t]{@{}l@{}}\textbf{FindTrendOutsideThres:} Which \\ machines have ever been outside of \\ the acceptable range?\end{tabular} & \begin{tabular}[t]{@{}l@{}}\textbf{LookupTrendThres:} Over the time\\ period, Machine X  has been:  above\\ the acceptable range; within the \\acceptable range; and/or below\\ the acceptable range.\end{tabular} \\ \bottomrule
\end{tabular}%
}

\label{tab:tasks}
\end{table}


Because range-based thresholds are prevalent in our identified application areas, we focus on thresholds bounded by an upper and lower value. 
Participants received additional framing for these tasks, providing context for an `acceptable range' of machine output. Our threshold tasks can be seen in \cref{tab:tasks}.

Our tasks can be situated in the context of visualization task taxonomies. In Brehmer and Munzner's typology~\cite{brehmer_multi-level_2013}, our tasks correspond to Discover $>$ Search $>$ Identify. Further, our \textbf{Lookup} tasks correspond to search tasks where both the target and location are known. Our \textbf{Find} tasks correspond to search where both the target and location are unknown (i.e., a search for entities based on their characteristics). In Lam et al.'s framework~\cite{lam2017bridging}, our tasks represent Explore and Describe goals for single populations. Their deeper levels of specificity (Explain and Confirm), such as identifying the main cause of an already identified problem, may occur in monitoring scenarios but are deeper follow-on analyses rather than rapid appraisal tasks that would be done with a glanceable display.

\subsection{Hypotheses}

Because of the complexity of some of our hypotheses, we include `simpler' hypotheses ahead of more complex ones. Many of these simpler hypotheses follow from existing literature on emphasis \cite{mairena_which_2022, mairena_baseline_2020}, visual channels \cite{munzner_visualization_2015, cleveland_graphical_1984, mackinlay_automating_1986}, and time series \cite{javed_graphical_2010, fuchs_evaluation_2013, brehmer_comparative_2020} but to test for the interaction between time series representation and current value emphasis, we declare them here as well. Additionally, our hypotheses are detailed for each experiment in order of priority, so that each hypothesis above applies to the hypothesis below. The simpler hypotheses provide necessary framing and context for the ones that follow and are intended to apply to those ones as well. For our hypotheses, performing better refers to both more correct responses and faster completion times unless otherwise specified.
We preregistered the hypotheses\footnote{\url{https://osf.io/wzf6a}}. The full text of our hypotheses and their rationale can be found in \cref{tab:hypotheses}.

\begin{table}[]
\caption{Experiment 1 and 2 Hypotheses and Rationale}

\resizebox{\columnwidth}{!}{%
\begin{tabular}{lll}
\toprule
 & \textit{Hypothesis} & \textit{Rationale} \\ \midrule
\multicolumn{3}{l}{\textbf{Experiment 1}} \\ \midrule
\textit{\textbf{H1A}} & \textbf{\begin{tabular}[c]{@{}l@{}}Test conditions (Emphasis, IntSize, \\ IntPos, SepSize, and SepPos) will \\ outperform the Control condition.\end{tabular}} & \begin{tabular}[c]{@{}l@{}}As expected from known principles \\ of visual encoding, any visual emphasis \\ should help people more rapidly scan \\ for current values. Results contradicting \\ H1A would be highly unexpected, but \\ this hypothesis was included as a basic \\ principle underlying subsequent hypotheses.\end{tabular} \\ \midrule
\textit{\textbf{H2A}} & \textbf{\begin{tabular}[c]{@{}l@{}}Integrated conditions (IntSize, \\ IntPos) will outperform the separated \\ conditions (SepSize, SepPos) for \\ current value tasks; this will not \\ affect trend tasks.\end{tabular}} & \begin{tabular}[c]{@{}l@{}}We believed integrated conditions would \\ have an advantage for current value tasks \\ because the information is dual-encoded\\ in a single location (with position on the \\ Y-axis). We had no specific reason to \\ expect that integrated encodings would\\ interfere with trend understanding, and \\ thus chose the null hypothesis for trends; \\ however, any contradiction to this null\\ hypothesis would be interesting.\end{tabular} \\ \midrule
\textit{\textbf{H3A}} & \textbf{\begin{tabular}[c]{@{}l@{}}Position conditions (IntPos, SepPos) \\ will perform better than the size\\ conditions (IntSize, SepSize) for\\ current value tasks; this will not\\ affect trend tasks.\end{tabular}} & \begin{tabular}[c]{@{}l@{}}For current value tasks, visual channel\\ theory suggests that position would be \\ more accurate than size. For trend tasks,\\ we had no specific reason to expect \\ any interference and thus opted for \\ the null hypothesis, but as with H2A, \\ any observed contradiction to the \\ null hypothesis would be interesting.\end{tabular} \\ \midrule
\textit{\textbf{H4A}} & \textbf{\begin{tabular}[c]{@{}l@{}}Size conditions (IntSize, SepSize) \\ will result in more incorrect responses.\end{tabular}} & \begin{tabular}[c]{@{}l@{}}We suspected that size may make some\\ tasks more difficult compared to \\ position or length. Large marks might \\ be visually distracting or make the \\ underlying position of the mark more \\ difficult to judge. Small marks may\\ not stand out sufficiently to serve as \\ visual emphasis and may make \\ labels harder to read.\end{tabular} \\ \midrule
\multicolumn{3}{l}{\textbf{Experiment 2}} \\ \midrule
\textit{\textbf{H1B}} & \textbf{\begin{tabular}[c]{@{}l@{}}Conditions using color to represent \\ thresholds (ControlColor, IntSizeColor, \\ SepSizeColor) will outperform band \\ representations (ControlBand, \\ IntSizeBand, and SepSizeBand).\end{tabular}} & \begin{tabular}[c]{@{}l@{}}We could not find existing empirical \\ evidence confirming which of these \\ common threshold encodings is\\ more effective in line charts. \\ Color is a prominent visual characteristic \\ that can draw visual attention, \\ so we expected a stronger \\ pop-out effect than with band.\end{tabular} \\ \midrule
\textit{\textbf{H2B}} & \textbf{\begin{tabular}[c]{@{}l@{}}Color integrated and separated \\ conditions (IntSizeColor, SepSizeColor) \\ will outperform ControlColor. \\ Band integrated and separated \\ conditions (IntSizeBand, SepSizeBand) \\ will outperform ControlBand.\end{tabular}} & \begin{tabular}[c]{@{}l@{}}We were most interested in how \\ integrated and separated conditions \\ would perform against each other, \\ but did not assume one would outperform\\  the other (except for H3B, below). \\ We expected current value encodings \\ to outperform control by drawing visual \\ attention to out-of-range current values.\end{tabular} \\ \midrule
\textit{\textbf{H3B}} & \textbf{\begin{tabular}[c]{@{}l@{}}SepBand will perform better on \\ current value tasks than IntSizeBand.\end{tabular}} & \begin{tabular}[c]{@{}l@{}}The rationale for H3B is that the \\ large size of the current value mark \\ could make it challenging to judge\\ whether the mark is within or outside \\ of the band in the integrated case. \\ By including additional reference bars\\ in relation to the separated, it may \\ improve performance compared to the\\ integrated design that may \\ occlude the threshold crossing. \end{tabular} \\ \bottomrule
\end{tabular}%
}
\label{tab:hypotheses}
\end{table}

\subsection{Stimuli}
\subsubsection{Rationale}

We chose sets of 16 entities for the small multiples,
a balance between too many plots (small plots, unrealistic for many domains) or too few (insufficient distractors).
Previous studies \cite{hosseinpour_examining_2024} found that more small multiples plots decreased accuracy for identification, comparison, and summarization.
While 16 plots may be low for some applications, it aligns with healthcare settings \cite{barry_needs_2026}.

Based on our design space, we include both integrated and separated current value representations. Integrated designs take advantage of embedding within the plot, redundantly encoded via the mark position; Separated designs offer a consistent place to look for the current value, and the size of the current value could potentially be much larger or cover a wider range. While we kept size scaling consistent between integrated and separated designs for this experiment, alternate layouts that allow for increasing the size of the separated current value could be explored in the future.

Experiment 1 results informed the design of our second experiment. We chose to exclude position due to its performance on \textbf{FindCurrValHigh} and \textbf{Low} and the lower ease-of-use rating (see \cref{sec-Exp1}). We decided to focus on size-based, redundantly-encoded emphasis, varying across the categories of our design space and two types of threshold indication: \textit{band}, where a shaded band is overlaid representing the range of the threshold, and \textit{color}, where data points outside the threshold range are colored differently than those inside. These two categories cover the intrinsic and extrinsic emphasis as described by Hall et al. \cite{hall_formalizing_2016}, where color encoding is intrinsic and the layered band is extrinsic.

\begin{figure} [t]
    \centering
    \includegraphics[width=\linewidth, alt={}]{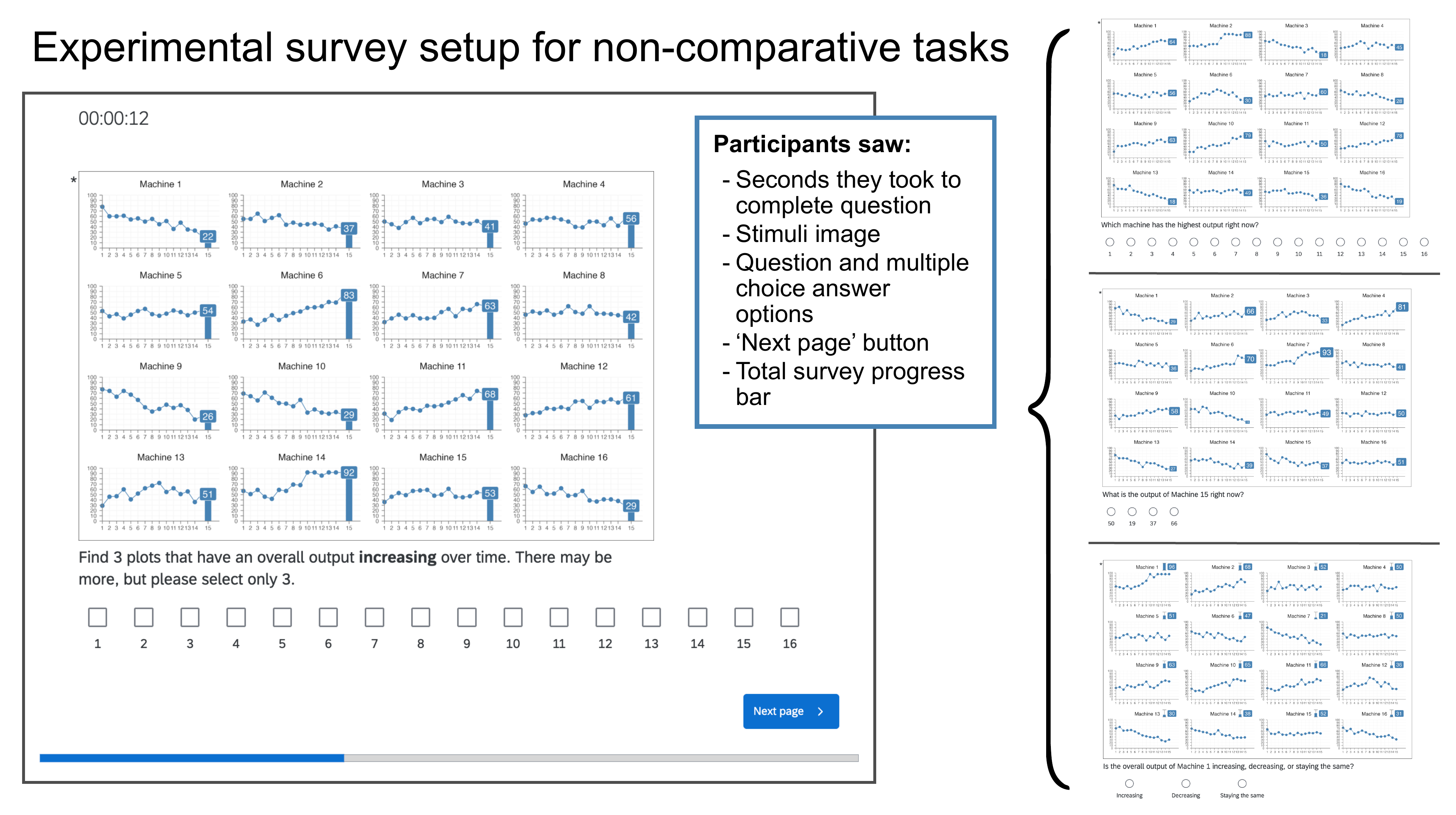 }
    \caption{Example survey pages showing what participants saw as they responded to each task. Question format had minor variations between tasks, and two similar questions asking for decreasing trends and lowest values are excluded here for brevity.}
    \label{fig:ExperimentalSetup}
\end{figure}

\subsubsection{Stimuli Design}
Across all tested designs, participants saw an image of a grid of 16 small plots, each representing the data output from one `machine' of many that they were monitoring. The machine output, in parts per minute, ranged from 1 to 100 on the Y-axis, and most machines maintained an output of approximately 50 parts per minute for the duration. Each plot showed time on the X-axis, ranging from 1, the oldest visible time, to 15, representing the current value. Plots used a line chart and overlaid point marks at each time interval. The presentation of the current value varied according to the design and redundant encoding (see \cref{fig:DesignVariations}).

Stimuli images were generated with synthetic data, following the same set of internal patterns with noise added. Data was generated in R \cite{rcoreteam_R_2024} and was seeded. For Experiment 1, each small multiple chart contained: two machines with a slightly increasing trend, two slightly decreasing, two strongly increasing, one exponentially increasing, three strongly decreasing, one with a slight peak in the middle, and five stable trends. Experiment 2 used the following trends for each small multiple chart: two machines with a slightly increasing trend, two slightly decreasing, one strongly increasing, one exponentially decreasing, peaking, or increasing above the upper threshold limit, one exponentially decreasing, peaking, or increasing below the lower threshold limit, one strongly decreasing, one with a slight peak in the middle, and eight stable trends. Each non-stable trend was randomly assigned to a number from 1 to 16 with no overlap, and stable trends were used for all remaining machines. All underlying trends were consistent across our designs and trials, so noise was generated from a normal distribution and generated for each frame for each image. Should the noise cause the point to exceed the 1 to 100 range, it was corrected within bounds; this resulted in `exponential' trends appearing to stabilize around the ranges of 91-100 or 1-10. For Experiment 2, the number of machines which exceeded the defined `acceptable range' of 30 to 70 parts per minute (inclusive) was restricted to four. If more than 4 machines were outside of this threshold for their current value, the closest to the threshold was adjusted to be inside the range. No ties between machines were permitted for the number of readings outside of the threshold. 

Images were created using \textit{ggplot2} \cite{wickham_ggplot2_2016}.
All stimuli had a labeled last point, and text size was held constant except where the stimuli varied in size; then, the text size varied with the label size over a set range of magnitude. Size scaling in \textit{ggplot2} maps a linear increase in a variable onto a linear increase in the area of the affected mark \cite{wickham_ggplot2_2016}. For Experiment 2, a threshold range was indicated by one of two methods: band or color. For the band, a light gray rectangle was underlaid on each time series chart, from 30 to 70 parts per minute. As needed, this band extended beyond the 1 to 15 range of the X-axis to include the full labeled point. For SepSizeBand, the band was underlaid across the frames, and an additional band was placed under, around, or above the label for the current value in the top right corner of each frame (shown in \cref{fig:SeparatedSizeBandExample}). In the color representation, points within 30 to 70 parts per minute were colored gray, points above 70 were orange, and points below 30 were blue. The chosen colors were of similar luminance. We manually reviewed the datasets generated for each task and selected a range of difficulties. For the \textbf{FindTrendOutsideThres} task for Experiment 2, we altered 3 datasets within our synthetic data to adjust the difficulty of the task. The code to generate the underlying data and the designs is available on OSF.

\subsection{Participants}
Participants were recruited online through Prolific and paid \$12 per hour. Participants used their own devices to complete the surveys but were required through Prolific to use a desktop or laptop device (not including tablets). Participants were U.S. users between the ages of 18 to 65 who are fluent in English and have a Prolific approval rating from 98 to 100. Participants who took part in any previous trials could not participate again. Both experiments took approximately 15 minutes. Participant demographics are detailed in the supplemental materials.

54 different participants completed each experiment. From pilot studies, we calculated effect sizes and conducted a power analysis using G\*Power to estimate sample size. We included additional participants in the case of exclusion following our preregistered protocol. We wanted to obtain 0.95 power in order to detect a 0.23 (Experiment 1) and 0.21 (Experiment 2) effect size at the standard 0.05 error, which gave us suggested participant groups of 33-39 people.

\subsection{Experimental Design and Procedure}

Experiment 1 used a 6 $\times$ 6 within-subjects study design, with 6 tasks (falling into 4 categories) and 6 design variations. For Experiment 1, we repeated the Find tasks for large and small values to isolate any impact caused by the size design variations. Experiment 2 used a 4 $\times$ 6 within-subjects study design, with 4 tasks and 6 design variations. Across both experiments, for each task, participants saw each design using one of 6 datasets. We used a 6 $\times$ 6 Latin square to control for which dataset appeared with which condition. 
In the Latin square design, participants were randomly assigned via the block randomizer in Qualtrics.
Trials were presented in a block design, so participants answered all questions for a given task before moving on; the order of trials within each block was randomized. For task questions that pertained to individual machines (Lookup tasks), the machine in question was randomly selected with quotas for types of trend, meaning no trend could be selected more than twice per block except for the stable trend, for a range of answer difficulty.

Experiments were conducted on Prolific, an online, crowd-sourced platform. Participants gave IRB-informed consent and confirmed they were using any necessary assistive technologies, 
e.g., reading glasses. Participants also confirmed they were completing the studies on a desktop or laptop computer with the browser window full screen. Following these checks, participants received a study overview, a framing scenario, instructions on how to read the control graph, and training questions. Training questions provided an explanation on how to read each design variation, which were labeled A-F in each experiment. For Experiment 1, participants completed a single type of task question across all design variations (FindCurrValHigh), and then trained on each other task question using the control design. In Experiment 2, participants completed two types of task questions (FindTrendOutsideThres and LookupCurrValThres) across all design variations and then trained on each other task question using the control band design. For both experiments, participants had to answer each training question correctly before moving on. 

Before beginning timed trials, participants were told to answer \textit{accurately} as quickly as possible but to prioritize accuracy over time. We included a pause page before each question so that participants could take breaks if needed. At full-screen size, no scrolling was necessary to read the question, see the visualization, answer, or click to the next page. A timer was displayed on each question page to remind participants they were being timed. In each experiment, three task types required selection of a single radio button, and one task type required the selection of three check boxes. Only three check boxes could be selected. An attention check asking for the number of machines on a sample control design variation was placed at the center of the experiment. 

Following the timed portion, participants were asked for feedback. They were shown a sample of each design (labeled A-F as found above   \cref{sec:methodology}) in training order and asked ``How easy was responding to tasks involving graphs with the above design (Design X)?'' We used a 5-point Likert scale with the following options: very easy, somewhat easy, neither easy nor difficult, somewhat difficult, and very difficult. We collected qualitative feedback using the prompt, ``Please compare the designs. Which ones were easiest to use to complete the tasks? Which were the worst? Why were they easier or worse?'' We also requested any additional feedback.
Finally, we collected basic demographic information and whether they had normal or corrected-to-normal vision. 

\section{Results}
We analyzed response time results using repeated measures ANOVA tests and mixed linear effect models for within-subjects designs using the R \cite{rcoreteam_R_2024} \textit{glmmTMB} package \cite{RJ-2017-066} to fit the models. For our accuracy analysis, we generated a binary result of \textit{0=incorrect} and \textit{1=correct} for each question based on a set of correct answers. To assess accuracy, we fit regularized binomial generalized linear mixed effects models, again using the \textit{glmmTMB} package \cite{RJ-2017-066}.

For time data, according to our pre-registered experimental design, we removed outliers by task for values outside of more than 1.5 times the interquartile range (IQR) beyond quartile 1 (below 25\% of responses) and quartile 3 (above 75\% of responses). We assessed the identified outliers for consistency across datasets and design variations and did not find them to be overly exclusive of any category. Response time outliers were not removed for accuracy analysis.
Timing data were natural log transformed following a check for normality. 
Following the transform, \textit{log(Time)} met at least one metric for normality for each task across experiments.

To test whether our hypotheses held when looking at response time, we used a repeated measures ANOVA for each task across all design variations. Specifically, we tested for differences between mixed linear effects models with random effects for participants and control models for each task, using the following formulas for our mixed effect models:

\vspace{-1.5em}
\begin{equation}
\label{eq:Equation1}
log(Time) = \beta_0 + \beta_1 \cdot DesignVariation + b_{Participant}
\end{equation}
\vspace{-11mm}

\begin{equation}
\label{eq:Equation2}
log(Time) = \beta_0 + b_{Participant}
\end{equation}

Equation 1 represents a linear mixed effects model, where the dependent variable is \textit{log(Time)}, $\beta_0$ is the intercept value, $\beta_1$ is the slope coefficient for $DesignVariation$, our main predictor, and $b_{Participant}$ is the random intercept used to control for participant response variation. 
Equation 2 uses the same variables but excludes the design variations term to act as a control case. If there is no significant difference that can be explained by the design variations, the models generated by Equations 1 and 2 will be similar. 

\subsection{Experiment 1 (Non-Threshold Tasks)}
\label{sec-Exp1}
Experiment 1 compared \textbf{Standard}, \textbf{Emphasis}, \textbf{Integrated}, and \textbf{Separated} conditions for non-threshold tasks (see Fig.~\ref{fig:DesignVariations} top). Five participants performed below the minimum number of correct responses ($\le$29 out of 36 questions) and were excluded. All participants answered the attention check correctly or answered correctly for the block the attention check was in, so 49 participant results were analyzed.

\begin{figure} [t]
    \centering
    \includegraphics[width=\linewidth, alt={Experiment 1 percent change from control means, confidence intervals, and error counts}]{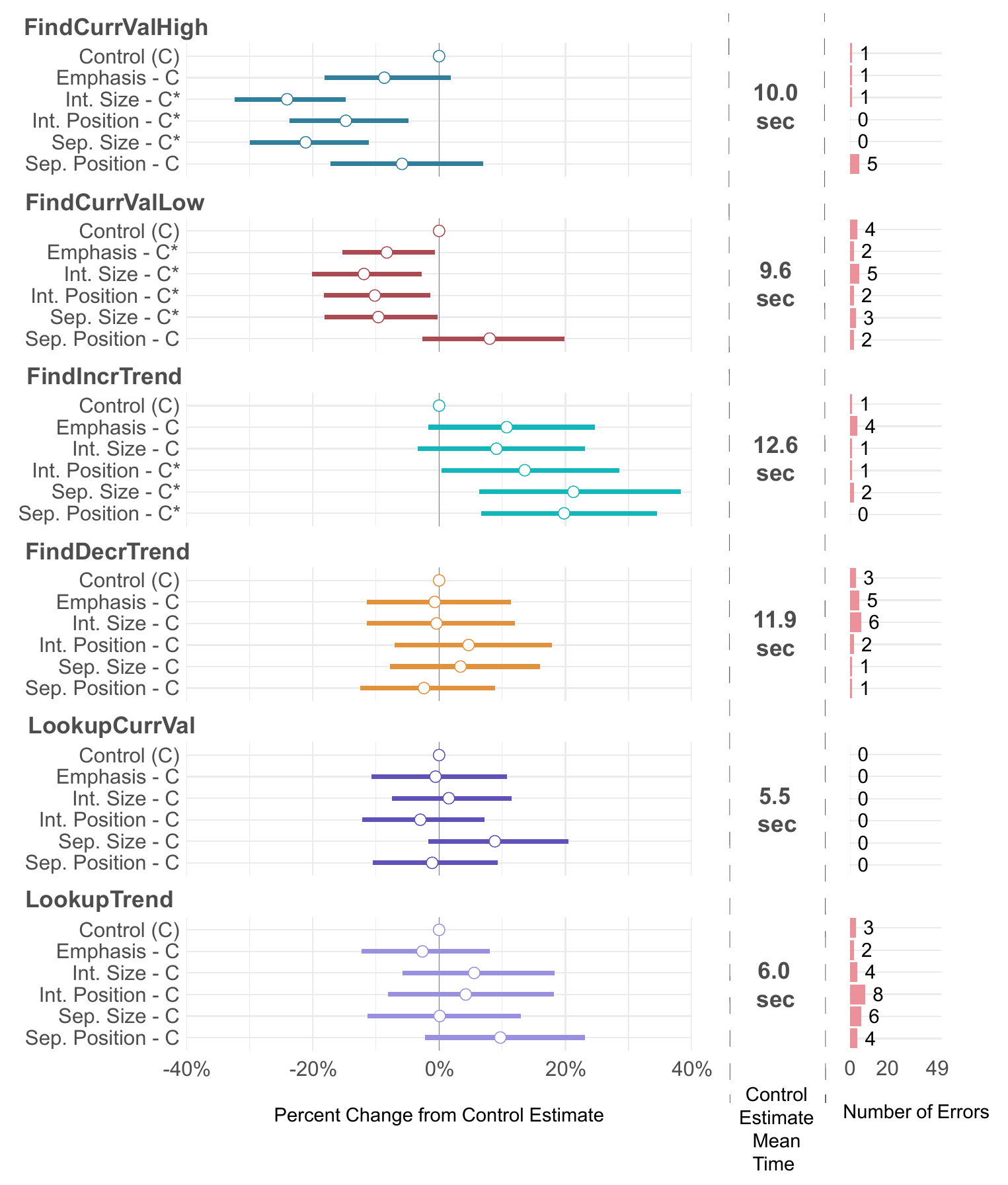}
    \caption{Exp. 1 response times as percent change from the control (C) estimated geometric mean time, where the bars represent the 95\% confidence intervals (left); Exp. 1 control estimated geometric mean times (center); Exp. 1 number of errors (right). Asterisks denote a significant difference from the control case for the estimated mean time.}
    \label{fig:nonComparativeCIs}
    \vspace{-5mm}
\end{figure}

To determine whether our design variations performed differently depending on the location of the information within the range of possible values, we tested both highest and lowest current value Find tasks and increasing and decreasing trend Find tasks. Overall percent change of model estimates and error results are shown in Fig.~\ref{fig:nonComparativeCIs}. We present our results using percent change from the control geometric mean time for each task and their scaled confidence intervals; visualizations showing raw means and confidence intervals as well as means normalized to the control mean by task are available in our supplemental materials. Because our models are fit on $log(Time)$, our estimates for mean time are computed using geometric means. Our results showed that most of the test conditions improved performance for finding highest and lowest current values but decreased performance at finding plots with an increasing trend.

For \textbf{FindCurrValHigh}, we found a significant ($p<.001$) overall difference between the control and design variations. Tukey post-hoc contrasts (done with the \textit{emmeans} package) showed a significant ($p<.001$) difference between Control and IntSize, SepSize and IntPos. According to our model's estimates, IntSize was 28\% faster than control; SepSize was 23\% faster. From our model's estimated control time of 10.0 seconds, this represents about a 2.5-second reduction in task time. IntSize and SepSize were also significantly different from Emphasis (Emphasis-IntSize: $p<.001$; Emphasis-SepSize: $p=.008$) and SepPos (SepPos-IntSize and SepPos-SepSize: $p<.001$ and $p=.003$). IntPos was significantly different from Control ($p<.001$) but not Emphasis. SepPos was not significantly different from the control. 

For \textbf{FindCurrValLow}, we found a significant ($p<.001$) overall difference.
From post-hoc testing, 
Emphasis, IntSize, IntPos, and SepSize all decreased the response time compared to Control by 9-13\%, with IntSize reducing the response time by the most (Control-Emphasis: $p=.025$; Control-IntSize: $p=.003$; Control-IntPos: $p=.014$; and Control-SepSize: $p=.042$).
SepPos performed worse than the control, resulting in a 7.7\% time increase; while SepPos's difference from Control was not significant, it was significantly ($p\leq.001$) different than all other design variations.

For \textbf{FindIncrTrend}, we found a significant ($p<.001$) overall difference between the control case and our design variations; however, this task resulted in an increase in time for our design variations. From post-hoc contrasts, we found a significant ($p<.001$) 19\% increase of SepSize over Control, or an increase of about 2.5 seconds per task. We also found significant differences between SepPos and Control (SepPos-Control: $p<.001$, 18\% increase past Control), between SepSize and Control (SepSize-Control: $p<.001$, 19\% increase past Control), and between IntPos and Control (IntPos-Control: $p=.04$, 13\% increase past Control). For \textbf{FindDecrTrend}, we did not find any significant overall difference ($p=.857$).

For both Lookup tasks, \textbf{LookupCurrVal} and \textbf{LookupTrend}, we found no overall significant differences (CurrVal: $p=.079$; Trend: $p=.138$). Between designs, we found significant differences between SepSize and both IntPos and SepPos for LookupCurrVal (SepSize-IntPos: $p=.008$; SepSize-SepPos: $p=.048$). For LookupTrend, we also found a significant difference between Empahsis and SepPos ($p=.04$), but no other contrasts were significant.

On accuracy, ANOVA tests found no significant differences between models incorporating our designs and models without. We found no significant differences in the distribution of errors across design variations for all other tasks. Most errors were in the \textbf{LookupTrend} task; in particular, for one `machine,' participants often interpreted a flat trend as decreasing. Since this difficulty was an artifact of one particular stimulus, we did not analyze it further.

\begin{figure} [t]
    \centering
    \includegraphics[width=\linewidth, alt={Experiment 1 Likert ratings showing participant preference for the design variations of integrated size and separated size, followed by emphasis and integrated position, followed by separated size and the control.}]{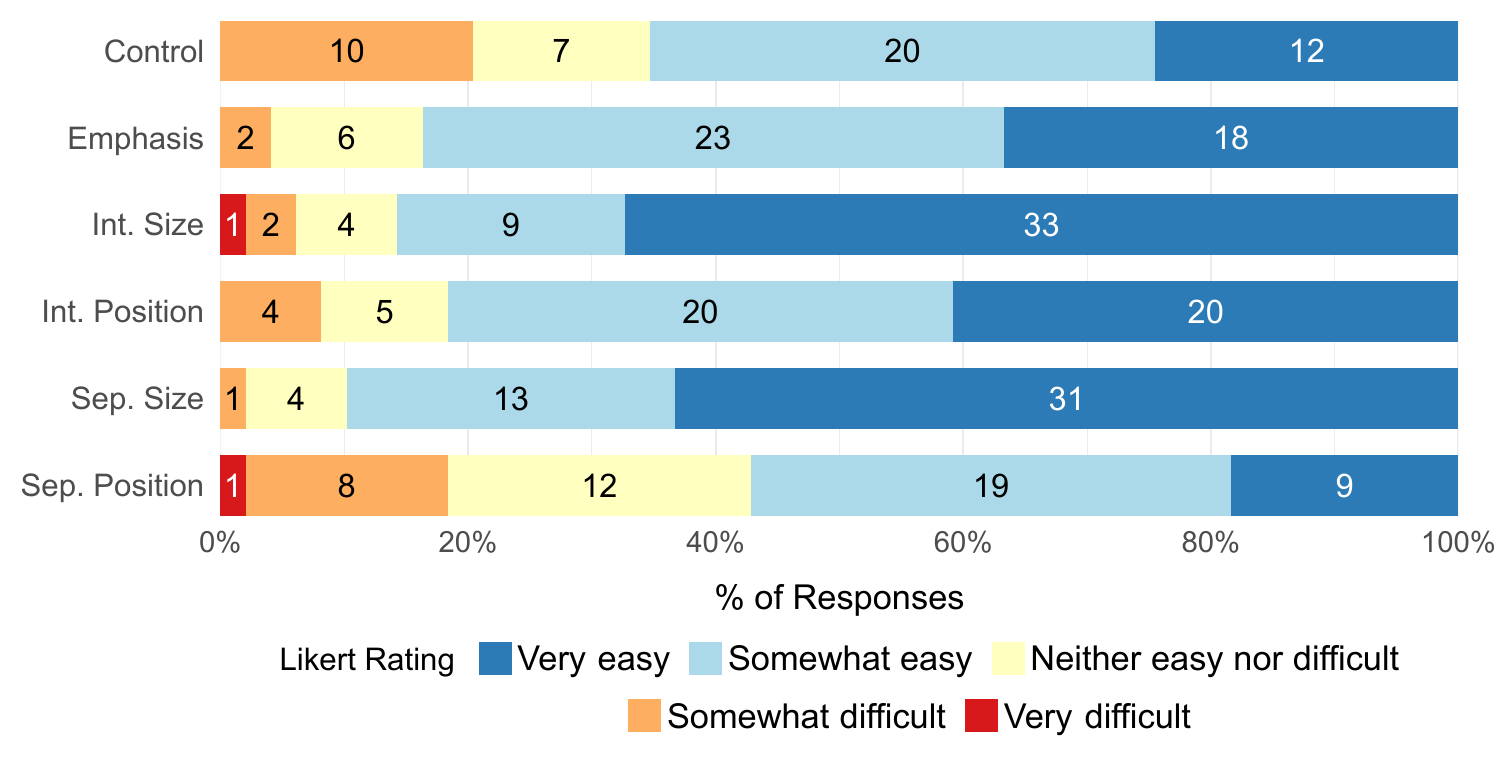}
    \caption{Exp. 1 Likert ratings. Bar labels show count of responses.}
    \label{fig:NoncomparativeLikert}
    \vspace{-2mm}
\end{figure}

Responses to our Likert scale rating formed pairs of ease of use, with IntSize and SepSize being most commonly rated as ``very easy'' to use, followed by Emphasis and IntPos, followed by Control and SepPos, as can be seen in \cref{fig:NoncomparativeLikert}. We found an overall significant difference ($p<.001$) between design variations from an ANOVA test. Additionally, these groupings were validated in post-hoc testing. IntSize ($p<.001$; $p<.001$) and SepSize ($p<.001$; $p<.001$) were significantly different than Control and SepPos but not each other ($p=0.999$). Emphasis ($p=0.025$; $p<.001$) and IntPos ($p=0.037$; $p=.0015$) were significantly different than Control and SepPos but not each other ($p=0.999$). IntSize and SepSize were not significantly different from Emphasis and IntPos (all $p>0.15$).
When asked why they found a design variation easier or more difficult, participant responses were mixed. Many participants liked the increased visual space occupied by the current value in the design variations, but several surfaced issues with size as an encoding, due not only to the increased space it occupies but the difficulty of reading the value when the size is decreased.

\subsubsection{Experiment 1 Discussion of Hypotheses}
Results of our hypotheses were mixed and depended on the task.
\textbf{H1A} (test conditions will outperform control) was confirmed for \textbf{FindCurrValue} tasks (both high and low), demonstrating that visually encoding the current value provided a benefit but only for these tasks. There were no significant improvements or detriments for \textbf{Lookup} tasks. Unfortunately, and contradicting \textbf{H1A}, the current value encodings interfered with the \textbf{FindIncrTrend} task compared to the control, and this was significant for the separated designs.

\textbf{H2A} (integrated will outperform separated for current value tasks) was confirmed for \textbf{FindCurrValue} tasks (high and low) but not for \textbf{LookupCurrValue} tasks. Interestingly, finding increasing trends was negatively impacted by our test conditions, particularly the separated ones. This contradicts \textbf{H2A}, which predicted no interference.

Hypothesis \textbf{H3A}, that position would outperform size, was generally unconfirmed by our results, as was Hypothesis \textbf{H4A}. In fact, size outperformed position on identifying the highest value. For our only significant result for accuracy, SepPos was the likely cause of the finding. 

\subsection{Experiment 2 (Threshold Tasks)}
Experiment 2 compared variations from our design space for threshold tasks, using band and color-based threshold representations (see Fig.~\ref{fig:DesignVariations} bottom). Three participants were excluded for performing below the minimum number of correct responses ($\le$15 out of 24 questions). Four additional participants were excluded for failing to correctly answer the attention check or answering incorrectly for the block the attention check was in. 47 participant results were analyzed.

\begin{figure}
    \centering
    \includegraphics[width=\linewidth, alt={Experiment 2 percent change from control means, confidence intervals, and error counts}]{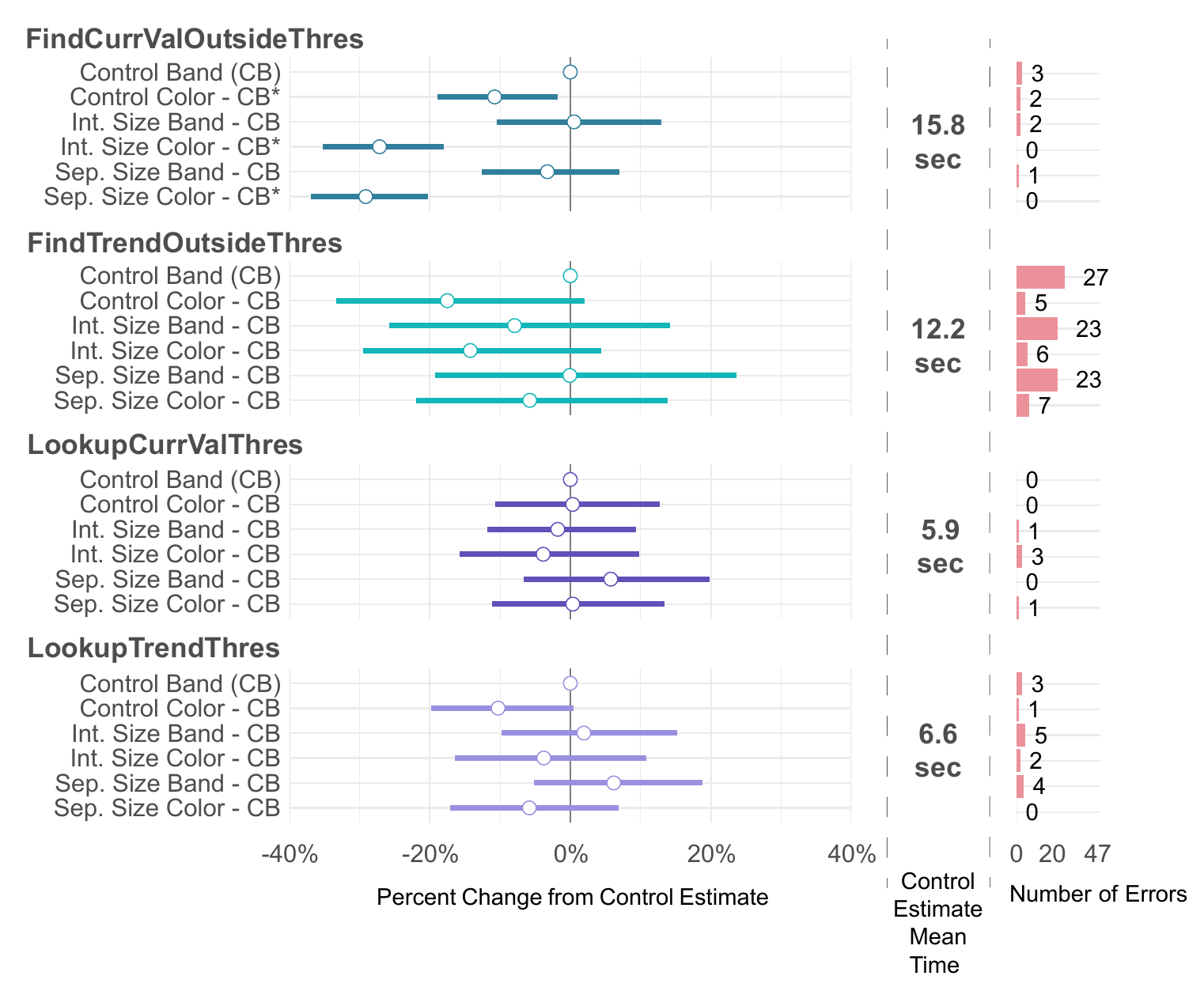}
    \caption{Exp. 2 response times as percent change from the control band (CB) estimated geometric mean time, where the bars represent the 95\% confidence intervals (left); Exp. 2 control band (CB) model-estimated geometric mean times (center); Exp. 2 number of errors (right).
    Asterisks denote a significant difference from control band.}
    \label{fig:ComparativeCIs}
    \vspace{-5mm}
\end{figure}

Overall model estimates and error results are shown in Fig.~\ref{fig:ComparativeCIs}; we again present our results using percent change from the model-estimated geometric mean, using ControlBand as our comparison for each task, and their confidence intervals. Results generally showed that color outperformed band for identifying plots with values outside the threshold range (FindCurrValOutsideThres and FindTrendOutsideThres tasks). Statistical details are below.

For \textbf{FindCurrValOutsideThres}, color outperformed band. We found a significant ( $p<.001$) overall difference between the control case and our design variations. Using post-hoc contrasts with the \textit{emmeans} package, we found significant ($p<.001$) differences between both IntSizeColor and SepSizeColor and all band designs and ControlColor. IntSizeColor was 32\% faster than ControlBand, and SepSizeColor was 34\% faster, which for an estimated mean task length of 15.8 seconds is a difference of approximately 4-5 seconds. ControlColor was also significantly different from ControlBand ($p=.009$; 11\% faster); however, it was not significantly different from the other band conditions, IntSizeBand ($p=.07$; 12\% faster), or SepSizeBand ($p=.29$; 8\% faster). IntSizeColor and SepSizeColor were not significantly different from each other ($p=.99$), with SepSizeColor 3\% faster than IntSizeColor. Band designs were also not significantly different from one another (all $p>.90$).

For \textbf{FindTrendOutsideThres}, we also saw a trend of times with color designs being faster than the corresponding band designs with a significant ($p=.03$) overall difference. However, post-hoc testing showed only one result of note: ControlColor was 19\% faster than SepSizeBand ($p=.054$), but this did not meet the criteria for a significant result. For \textbf{LookupCurrValThres}, we found no significant differences ($p=.31$). For \textbf{LookupTrendThres}, we found a significant ($p=.013$) overall difference but only two significant pairwise differences: ControlColor was 13\% faster than IntSizeBand ($p=.032$) and 17\% faster than SepSizeBand ($p<.001$).

On accuracy, we found no significant differences between our design variations and the control case for \textbf{FindCurrValOutsideThres}, \textbf{LookupCurrValThres}, or \textbf{Lookup Trend}. However, we did find a significant ($p<.001$) overall difference for \textbf{FindTrendOutsideThres}. ControlColor, IntSizeColor, and SepSizeColor were all significant ($p<.01$) in the overall model and were significant in post-hoc testing against all band conditions. All color variations were not significant with each other, and all band variations were not significant with each other. Many errors came from three datasets within the \textbf{FindTrendOutsideThres} condition. These datasets all had trends that were outside of the threshold range at both the start and at the end of the time frame, leading to an overall greater time spent outside of the threshold. With the band representation of these datasets, participants more frequently chose a machine that had the longest \textit{\textbf{contiguous}} out-of-range time rather than adding up the total of these split time frames (see \cref{sec:Exp2Discussion} and \cref{fig:DemonstrateErrors}).

\begin{figure} 
    \centering
    \includegraphics[width=\linewidth, alt={Experiment 2 Likert ratings showing participant preference for the design variations using color over band threshold design.}]{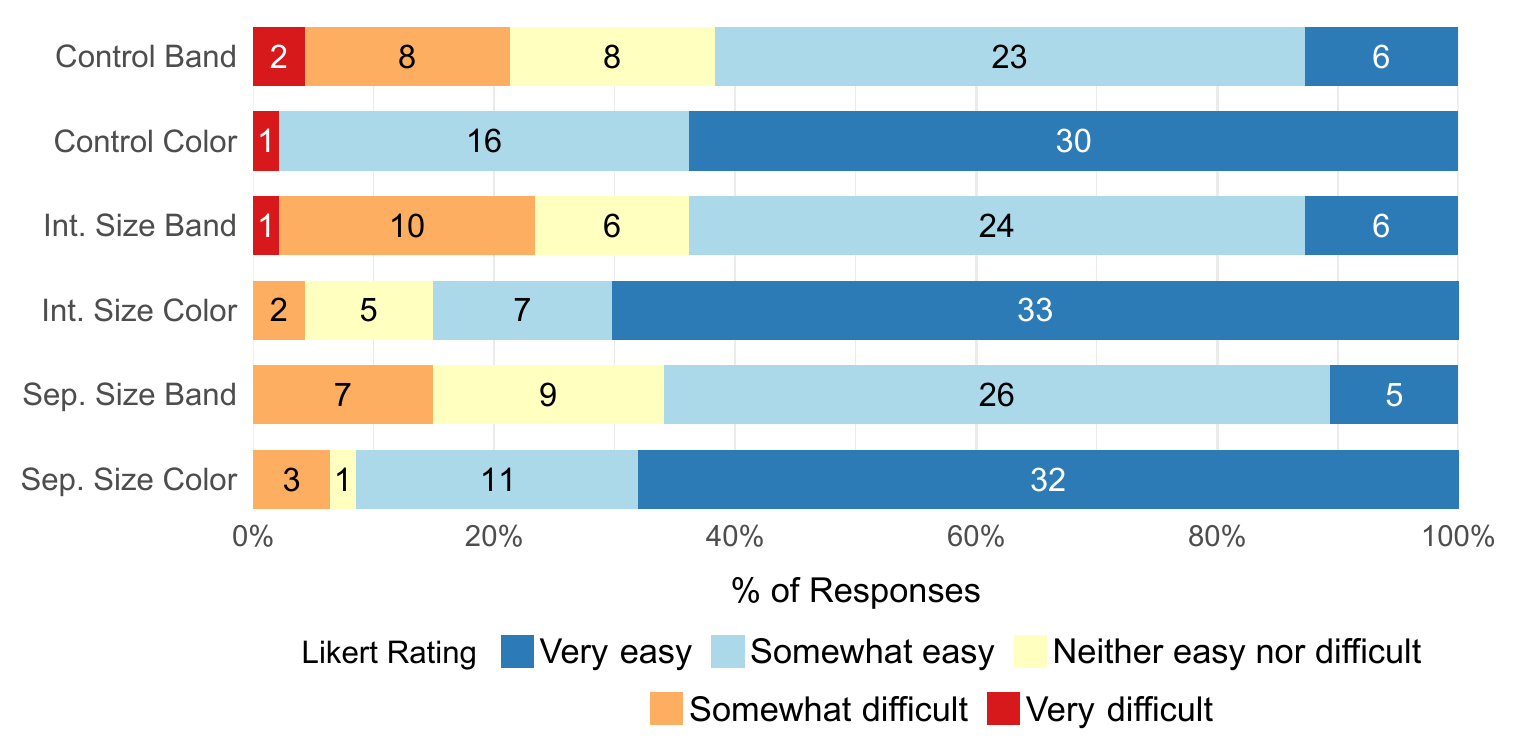}
    \caption{Exp. 2 Likert ratings. Bar labels show count of responses.}
    \label{fig:ComparativeLikert}
    \vspace{-5mm}
\end{figure}

Likert responses on ease of use favored color over band for all design variations (see \cref{fig:ComparativeLikert}). We found an overall significant ($p<.001$) difference; post-hoc tests showed that all color variations were significantly (all: $p<.001$) different than all band variations. There were no significant differences among band variations or among color variations. Interestingly, more participants rated ControlColor as very easy or somewhat easy to use than IntSizeColor or SepSizeColor. Participants consistently rated band conditions as more difficult.

From qualitative responses, participants generally preferred the color variations over band. Some also preferred the larger labels over the control labels. Participants were mixed in preference between IntSizeColor and SepSizeColor. A small subset preferred SepSizeBand to quickly assess the current value's relation to the threshold, though other participants found this design the hardest to use because of the amount of gray shading present in the design (see \cref{fig:SeparatedSizeBandExample}). Several participants suggested combining the band and color representations.

\subsubsection{Experiment 2 Discussion of Hypotheses}
Similarly to Experiment 1, the results of our hypotheses were mixed and varied across tasks. Hypothesis \textbf{H1B}, that color would outperform band, was strongly confirmed, though the difference primarily applied to \textbf{Find} tasks, not to \textbf{Lookup} tasks.

Hypothesis \textbf{H2B}, that integrated and separated designs would perform better than their respective controls, was confirmed for color designs on the \textbf{FindCurrValOutsideThres} task, demonstrating the value of current value encodings when assessing current status. Other tasks and designs were not found significantly different than their respective controls. Importantly, none of the encodings significantly degraded performance compared to the control on any task.

Hypothesis \textbf{H3B}, that SepSizeBand would perform better on current value tasks than IntSizeBand was unconfirmed, though SepSizeBand did perform the best of the band designs for the \textbf{FindCurrValOutsideThres} task.

\section{Discussion}
\subsection{Experiment 1 (Non-Threshold Tasks)}
In Experiment 1, we found consistent results across \textbf{FindCurrVal} and \textbf{Lookup} tasks, but inconsistent results between our \textbf{FindIncrTrend} and \textbf{FindDecrTrend} tasks, which was surprising. All participants received the task sets in the same order as we do not compare across tasks, reducing the need for counterbalancing, but the effect is larger than we would expect to see for learning the question. We also included training to limit learning effects. Perhaps participants were distracted by the current value representation or found it harder to identify the trend due to interference from the emphasized current value. 
With decreasing trends, shorter bars and smaller labels may have been less obstructing than for increasing trends. Further study is needed to identify the cause of this effect, potentially through an eye-tracking experiment.

Our intuition that position would perform better than size (Hypothesis H3A) was generally not confirmed. While this does not align with previous research on visual encodings \cite{munzner_visualization_2015, cleveland_graphical_1984} as we had expected, it does align with literature on emphasis design \cite{mairena_which_2022, mairena_baseline_2020}, where size has been found to be particularly effective. We had expected that position encodings may perform better in comparison while size encodings may be more variable, such as being difficult to differentiate at smaller sizes. However, our separated and integrated position conditions performed differently, with separated position performing worst overall. In aggregate, any interference effects that may exist from a size encoding of the current value were not substantial enough in this experiment to impact the performance of the integrated size condition. This suggests that the distinction between separated and integrated designs can make an impact on how encodings are perceived.

We also found little variation in results across our design variations for our \textbf{FindDecrTrend}, \textbf{LookupCurrVal}, and \textbf{LookupTrend} tasks, leaving our hypotheses mostly unconfirmed for these tasks. In particular, we found that our emphasis techniques did not significantly hinder trend or current value identification of individual plots (\textbf{Lookup} tasks) or across decreasing small multiples (\textbf{FindDecrTrend}). This is, in fact, a very promising result, because it suggests that encodings that assist with assessment of the current status have little interference with other tasks, though further study could add to this evidence with other design variations and channels of emphasis. There was, however, some interference with \textbf{FindIncrTrend}. Of the designs we studied, integrated designs suffered less interference than separated ones.

In the Likert ratings and qualitative responses, participants preferred redundantly-encoded presentations compared to the control and emphasis. This could inform future work about emphasis via repetition of information through additional channels, which may contribute differently to visual prominence than a flat application of a channel.

In general, we found that the best designs for finding current values and trends likely depends on the user's predominant tasks. For tasks focused on current values, size-based techniques generally performed well, but position encoding, particularly when separated, was less effective. For trend tasks, additional visual weight for the current value did seem to impact trend assessment, particularly increasing trends if size increases with the value. Separated positional encodings were less effective across multiple tasks, or no more effective than other methods, making them likely a poor overall choice. Integrated size perhaps offers the best overall compromise, especially in use cases where finding high or low current values is frequent, with up to 28\% model-estimated mean response time benefit for Find Current Value tasks and no significant difference from control for Find Increasing Trend tasks, unlike our other Integrated and Separated designs.

\subsection{Experiment 2 (Threshold Tasks)}
\label{sec:Exp2Discussion}

\begin{figure}
    \centering
    \includegraphics[width=\linewidth, alt={For threshold designs, errors are more easily identifiable through color encodings, in part because errors split between different areas of a trend are more visible}]{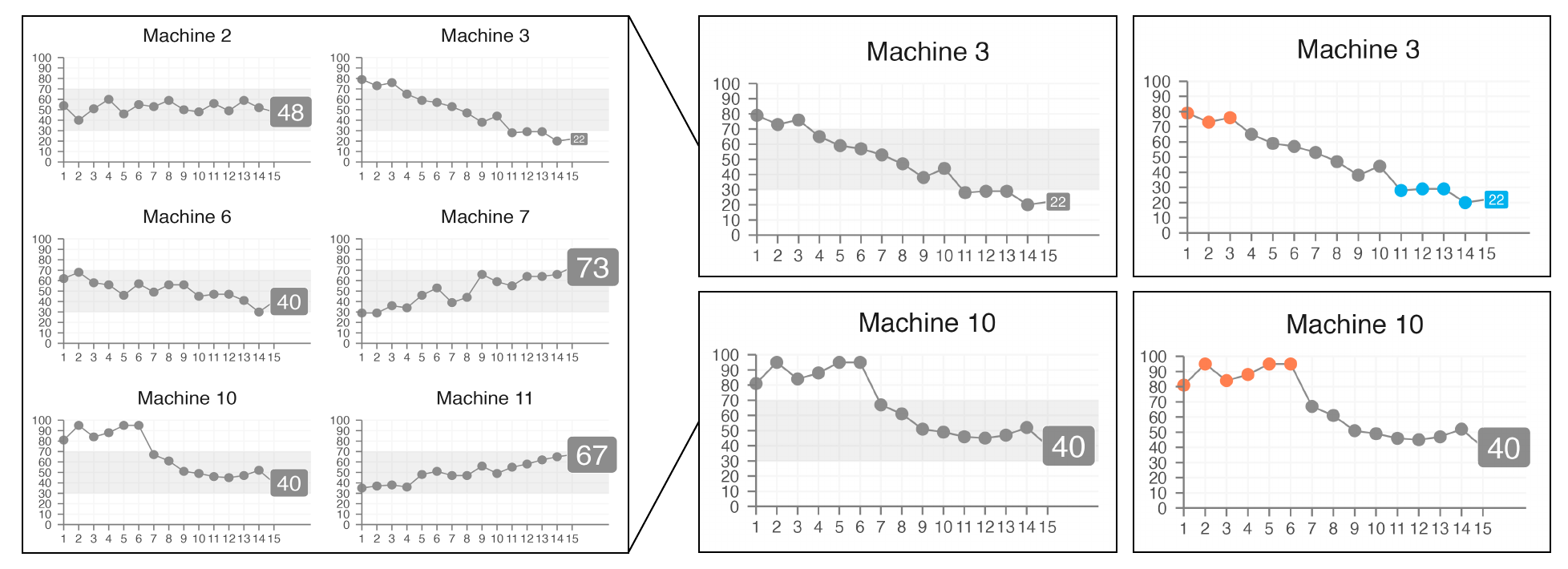}
    \caption{Threshold deviations are more visible in color designs than band designs, particularly when the values are out of range throughout the time series (top) rather than contiguously (bottom).}
    \label{fig:DemonstrateErrors}
\end{figure}

We found consistent evidence in favor of color threshold designs. 
For \textbf{Find} and \textbf{LookupTrendThres} tasks, performance was faster with color than with band; \textbf{FindCurrValOutsideThres} had significant improvement in time and \textbf{FindTrendOutsideThres} in error rates, though the trends in \textbf{LookupTrendThres} were not significant. Color is pre-attentive \cite{munzner_visualization_2015}, and in this usage, has the added benefit of being able to differentiate points on the edge of a fine threshold barrier much more clearly. On accuracy, color dramatically outperformed band on \textbf{FindTrendOutsideThres} tasks. In particular, band designs were worse for finding trends where points were outside of the threshold in multiple places in the time range rather than contiguous (see \cref{fig:DemonstrateErrors}). Participants also had more errors for band conditions on \textbf{LookupTrendThres} tasks, but these error rates were less substantial.

Results were mixed for integrated and separated designs. Both colored variants performed better with some designs than the controls for \textbf{FindCurrValOutsideThres}. However, results were less clear for other tasks. Integrated designs performed better in aggregate than their respective separated counterparts for \textbf{FindTrendOutsideThres}, though not significantly, and the control cases, particularly for color, performed on par with the integrated and separated design variations. Similar to Experiment 1, there was also no variation between integrated and separated design performance and control performance for \textbf{LookupTrendThres} and \textbf{LookupCurrValThres}.

The emphasis placed through color for the \textbf{FindCurrValOutsideThres} task is a clear result, with increased amounts of color on the current value helping participants to answer these questions 5 seconds (32-34\%) faster on average than ControlBand and 3-3.5 seconds (20-23\%) faster  than ControlColor, suggesting that the visual weight given by the increased size helped participants answer faster.

We had suspected that the unique design of the label for our separated band condition may help it perform better than the integrated band condition. Given the size of the label for the integrated band, we hypothesized that it may interfere with identifying if the current value was inside or outside the band. While the separated band condition did perform the best of the band conditions for \textbf{FindCurrValOutsideThres}, this was not a significant difference. For \textbf{LookupCurrValThres}, the design had no effect, which was surprising. As discussed above, the time to identify a given plot within the small multiples grid is likely the most impactful factor for this task, not the assessment of the current value relative to the threshold.

On average, color designs outperformed band designs on all tasks except \textbf{LookupCurrValThres}. IntSizeBand also performed similarly to SepSizeColor on the \textbf{FindTrendOutsideThres} task. Thus, we believe that color is more effective for indicating thresholds in dense scenarios like small multiples. If a threshold is crucial in a data monitoring scenario, color could be reserved for indicating a threshold instead of categories or other variables of interest.
Future work could explore combined color plus band designs as suggested by our participants, but band alone is not recommended based on our results.

\subsection{General Discussion}
Across our experiments, we found the most substantial results on \textbf{Find} tasks, while differences in \textbf{Lookup} tasks were generally not statistically significant.
Because \textbf{Lookup} tasks were focused on a single `machine' within a set of 16, emphasis techniques likely had less impact on assessing the trend and current value. For tasks focused on the whole set of small multiples (\textbf{Find} tasks), emphasis and redundant encodings generally benefited current value tasks, with smaller improvements or declines for trend tasks, potentially due to visual interference. In general, these results run counter to those of Chun et. al \cite{chun_redundant_2017}, where redundant encodings did not meaningfully impact performance. Our results are more consistent with Nothelfer et. al \cite{nothelfer_redundant_2017}, which point to redundant encodings of color and shape aiding in viewer identification. Here, we find tasks related to the individual plots not to be strongly impacted by our redundant encodings. However, tasks that involve looking across multiple plots were impacted by our design variations, including using color, so we recommend weighing the predominant task when making design decisions.

Separated designs performed generally less well than integrated. 
In Experiment 1, this was generally attributable to separated position, which was ineffective. In Experiment 2, performance was more mixed: the unusual separated band design did perform the best of the band designs for the \textbf{FindCurrValOutsideThres} task, but this result was not significant. Future research could tease apart the conditions under which separated designs perform effectively.
While we cannot generalize beyond the rapid appraisal tasks and design variations that we studied, our results suggest that integrated designs may be a promising overall design direction, whereas separated displays may perform well in more specific conditions.

For glanceable monitoring displays where contextual trend information is critical, we recommend balancing between the representation of the current value and the historical trend. Separated Position, at least in our instantiation, was not very effective and is not recommended. For non-threshold tasks, both Separated and Integrated Size, plus Integrated Position, assisted with current value tasks but all interfered with trend tasks to some degree. \textit{Integrated designs had the least interference among the tasks and displays we studied} and may offer a good balance. \textit{For threshold tasks in our experiment, color was a clear winner over band designs}. Both Integrated and Separated Size encodings performed well in this context.

For visualizations using size as an emphasis technique, there is a limit to the magnitude of size difference that can be accomplished before other data is obscured or the historical trend representation becomes unclear. It is possible that the benefit of size encoding may drop off with an increasing number of plots. Color might also become less effective at scale, as the size of the colored marks shrinks. Limiting the number of entities in a small multiple view not only aids in performance \cite{hosseinpour_examining_2024} but also potentially in the effectiveness of emphasis encoding. Nonetheless, applications may require more entities in the small multiple visualization than we included here. In these cases, size may become a difficult modality to use, and color may be more effective. Further research is needed to assess 
a larger (or smaller) number of multiples. Hosseinpour et al. \cite{hosseinpour_examining_2024} highlight specific entities via their border, which may work at high densities. If possible, designers could also break up visualizations into logical groupings across multiple monitors. 

\subsection{Limitations and Future Work}

Future work replicating our studies with eye-tracking may help explain the discrepancy in response times between the \textbf{FindIncrTrend} and \textbf{FindDecrTrend} tasks. Additionally, 
future work could empirically test different trend patterns and additional variations of emphasis, redundant encoding, and threshold design, including redundantly encoding thresholds (e.g., combining color and band). 
Additional work could test different trend types. 

The controlled nature of our studies allowed us to investigate broadly applicable design concepts of \textit{separation} and \textit{integration}, and their interaction with various \textit{current value emphasis} and \textit{threshold} encodings, but we did not test all possible encoding choices.
Future work could test various conditions held fixed in these experiments, such as the number and arrangement of plots, number of data measures, length of the historical trend, and scale and variability of the data. 
Designs with wider aspect ratios may allow features of the separated designs to be more prominent or larger, which may impact their performance. For experimental tractability, we also did not test a separated variation with a ``big number'' for the current value, a common approach in medical monitors. 
Future work could also review polar coordinates and other scaling methods, as well as 
alternate types of visualizations 
such as bar charts, strip plots, comet plots, and others.

Due to the nature of an online study, we cannot guarantee that all participants followed instructions regarding window sizing and using any needed assistive devices in spite of confirming they had done so. Future work in-person could validate the results found here.

Our research investigated rapid appraisal tasks that are critical low-level components of many data monitoring activities. For experimental simplicity, we simplified trend data to increasing, flat, and decreasing linear trends. Data monitoring may involve tasks not considered here, including anomaly detection, more complex trend characterization, entity comparison, delta-checking, multi-factor tasks, situational awareness, and root cause analysis~\cite{cobilean2023review}. Future work is needed to expand our studies to these scenarios. Nonetheless, our low-level Find and Lookup tasks are likely a component of more complex monitoring activities. Future work is needed to expand our studies to other rapid appraisal tasks (specifically, anomaly detection, trend characterization, and summarization for situational awareness), to datasets with more complex trends, and to multi-factor scenarios. 
Future research can also explore 
field research in real monitoring environments.

Monitoring scenarios typically involve dynamically-updating displays. For tractability and precise experimental control, our experiments here used static displays, which stand in as slowly-updating displays (e.g., on a timescale of minutes). Given our most common use cases, the use of static displays is the biggest limitation of this work. It will be important for future work to conduct similar experiments with dynamically-updating displays, including of varying update frequency. Dynamic displays would also allow animated methods of emphasis to be incorporated, something we could not test with static displays. Using this work as a foundation, the authors are conducting follow-up work in dynamic environments to validate the results found here for rapid streaming data use cases. Moreover, data monitoring tasks are typically performed by domain expert operators. We recommend future work comparing performance of trained monitoring experts to the general population.

Additionally, many of the monitoring scenarios we considered are examples of situated visualization, which locate data visualizations near a physical referent such as a patient or a place \cite{willett_embedded_2017}. Experience with situated visualizations for data monitoring has demonstrated the importance of glanceable displays due to the secondary nature of monitoring and the interruptions that occur in distracting environments \cite{vermeulen_informing_2012, heyer_investigations_2010}. We intend to conduct future work to address monitoring in these scenarios, using a dual-task paradigm.

Designing our experiments also uncovered the need for visualization research to develop a comprehensive framework for rapid appraisal data monitoring tasks. Our characterization of Lookup and Find tasks across trends and current values, with and without thresholds, is a start to a taxonomy. A more comprehensive framework would benefit future design and experiments for visualization-supported data monitoring.

\section{Conclusion}
Time series small multiples have complex design needs when users need to rapidly assess both the current value and the historical trend of many datasets at once. In this paper, we contributed a novel design space of solutions and a vocabulary to discuss and explore the problem of integrating current and historical values in individual charts. We also contributed two studies that provide initial evidence of the advantage of different design choices selected from the design space. In our experiments, we found that a combination of emphasis design principles made the current value more identifiable from a historical trend but that these visual encodings could sometimes interfere in the identification of trends for facets in the small multiple display. Integrated current value designs had lower overall interference than separated ones for the Lookup \& Find tasks and design variations that we examined. We also found that color-based methods of threshold design significantly outperformed band-based designs and that size-based current value encodings had minimal to no interference with threshold tasks.
We conclude that in order to design effective time series small multiples, methods of emphasis must be carefully weighed based on the most common usage of the visualization, how dense the visualization is, and whether thresholds and limits on the data need to be monitored. Overall, we suggest that designers tasked with presenting current values and trends simultaneously prioritize integrated designs and size encoding (while being mindful of its potential interference effect) and use color rather than bands for thresholding.

\section*{Supplemental Materials}
\label{sec:supplemental_materials}

All supplemental materials are available at \url{https://osf.io/wzf6a}.
An additional copy of these supplemental materials has been included with this paper.

\section*{Acknowledgments}
The authors wish to thank the Khoury Vis Lab at Northeastern University, and in particular, we would like to thank Lace Padilla for her support during the process. This work was supported in part by the Roux Institute.

\bibliographystyle{IEEEtran}
\bibliography{BibTimeSeriesSmallMutliples}

@article{cobilean2023review,
  title={A review of visualization methods for cyber-physical security: Smart grid case study},
  author={Cobilean, Victor and Mavikumbure, Harindra S and Mcbride, Brady J and Vaagensmith, Bjorn and Singh, Vivek Kumar and Li, Ruixuan and Rieger, Craig and Manic, Milos},
  journal={IEEE Access},
  volume={11},
  pages={59788--59803},
  year={2023},
  publisher={IEEE}
}

@article{lam2017bridging,
  title={Bridging from goals to tasks with design study analysis reports},
  author={Lam, Heidi and Tory, Melanie and Munzner, Tamara},
  journal={IEEE TVCG},
  volume={24},
  number={1},
  pages={435--445},
  year={2017},
  publisher={IEEE}
}

@inproceedings{nowell_graphical_2002,
	title = {Graphical encoding for information visualization: an empirical study},
	shorttitle = {Graphical encoding for information visualization},
	url = {https://ieeexplore.ieee.org/document/1173146/},
	doi = {10.1109/INFVIS.2002.1173146},
	urldate = {2025-02-06},
	booktitle = {{INFOVIS} 2002.},
	author = {Nowell, L. and Schulman, R. and Hix, D.},
	month = oct,
	year = {2002},
	note = {ISSN: 1522-404X},
	pages = {43--50},
}

@article{quadri_survey_2022,
	title = {A {Survey} of {Perception}-{Based} {Visualization} {Studies} by {Task}},
	volume = {28},
	issn = {1941-0506},
	url = {https://ieeexplore.ieee.org/document/9492011/?arnumber=9492011},
	doi = {10.1109/TVCG.2021.3098240},
	number = {12},
	urldate = {2025-02-06},
	journal = {IEEE TVCG},
	author = {Quadri, Ghulam Jilani and Rosen, Paul},
	month = dec,
	year = {2022},
	pages = {5026--5048},
}

@inproceedings{franke_comparative_2022,
	title = {A {Comparative} {Study} of {Visualizations} for {Multiple} {Time} {Series}:},
	isbn = {978-989-758-555-5},
	shorttitle = {A {Comparative} {Study} of {Visualizations} for {Multiple} {Time} {Series}},
	url = {https://www.scitepress.org/DigitalLibrary/Link.aspx?doi=10.5220/0010761700003124},
	doi = {10.5220/0010761700003124},
	language = {en},
	urldate = {2025-02-06},
	booktitle = {Proceedings of VISIGRAPP'22},
	author = {Franke, Max and Knabben, Moritz and Lang, Julian and Koch, Steffen and Blascheck, Tanja},
	year = {2022},
	pages = {103--112},
}

@book{aigner_visualization_2023,
	address = {London},
	title = {Visualization of {Time}-{Oriented} {Data}},
	copyright = {https://creativecommons.org/licenses/by/4.0},
	isbn = {978-1-4471-7526-1 978-1-4471-7527-8},
	url = {https://link.springer.com/10.1007/978-1-4471-7527-8},
	language = {en},
	urldate = {2025-02-06},
	publisher = {Springer},
	author = {Aigner, Wolfgang and Miksch, Silvia and Schumann, Heidrun and Tominski, Christian},
	year = {2023},
	doi = {10.1007/978-1-4471-7527-8},
}

@article{scheer_visualization_2022,
	title = {Visualization {Techniques} of {Time}-{Oriented} {Data} for the {Comparison} of {Single} {Patients} {With} {Multiple} {Patients} or {Cohorts}: {Scoping} {Review}},
	volume = {24},
	shorttitle = {Visualization {Techniques} of {Time}-{Oriented} {Data} for the {Comparison} of {Single} {Patients} {With} {Multiple} {Patients} or {Cohorts}},
	url = {https://www.jmir.org/2022/10/e38041},
	doi = {10.2196/38041},
	language = {EN},
	number = {10},
	urldate = {2025-02-06},
	journal = {JMIR},
	author = {Scheer, Jan and Volkert, Alisa and Brich, Nicolas and Weinert, Lina and Santhanam, Nandhini and Krone, Michael and Ganslandt, Thomas and Boeker, Martin and Nagel, Till},
	month = oct,
	year = {2022},
	pages = {e38041},
}

@article{chun_redundant_2017,
	title = {Redundant {Encoding} in {Data} {Visualizations}: {Assessing} {Perceptual} {Accuracy} and {Speed}},
	volume = {24},
	issn = {1555-1393},
	shorttitle = {Redundant {Encoding} in {Data} {Visualizations}},
	url = {https://www.tandfonline.com/doi/full/10.1080/15551393.2017.1343153},
	doi = {10.1080/15551393.2017.1343153},
	number = {3},
	urldate = {2025-02-05},
	journal = {Visual Communication Quarterly},
	author = {Chun, Russell},
	month = jul,
	year = {2017},
	pages = {135--148},
}

@article{nothelfer_redundant_2017,
	title = {Redundant encoding strengthens segmentation and grouping in visual displays of data.},
	volume = {43},
	copyright = {http://www.apa.org/pubs/journals/resources/open-access.aspx},
	issn = {1939-1277, 0096-1523},
	url = {https://doi.apa.org/doi/10.1037/xhp0000314},
	doi = {10.1037/xhp0000314},
	language = {en},
	number = {9},
	urldate = {2025-02-05},
	journal = {Journal of Experimental Psychology: Human Perception and Performance},
	author = {Nothelfer, Christine and Gleicher, Michael and Franconeri, Steven},
	month = sep,
	year = {2017},
	pages = {1667--1676},
}

@article{borgo_empirical_2012,
	title = {An {Empirical} {Study} on {Using} {Visual} {Embellishments} in {Visualization}},
	volume = {18},
	issn = {1941-0506},
	url = {https://ieeexplore.ieee.org/document/6327282/?arnumber=6327282},
	doi = {10.1109/TVCG.2012.197},
	number = {12},
	urldate = {2025-02-05},
	journal = {IEEE TVCG},
	author = {Borgo, Rita and Abdul-Rahman, Alfie and Mohamed, Farhan and Grant, Philip W. and Reppa, Irene and Floridi, Luciano and Chen, Min},
	month = dec,
	year = {2012},
	pages = {2759--2768},
}

@article{simpson2019false,
    title = {False alarms and overmonitoring: major factors in alarm fatigue among labor nurses},
    author = {Simpson, Kathleen Rice and Lyndon, Audrey},
    journal = {J.~Nurs.~Care Qual.},
    volume = {34},
    number = {1},
    pages = {66--72},
    year = {2019},
    publisher = {LWW},
    doi = {10.1097/NCQ.0000000000000335},
}

@article{han2007designing,
  title={Designing a human--computer interface for a process control room: A case study of a steel manufacturing company},
  author={Han, Sung H and Yang, Huichul and Im, Dong-Gwan},
  journal={International Journal of Industrial Ergonomics},
  volume={37},
  number={5},
  pages={383--393},
  year={2007},
  publisher={Elsevier}
}

@article{betancur2022visualization,
  title={Visualization proposal for power system control rooms based on situational awareness},
  author={Betancur, Daniel and Escobar, Eros David and Quintero, Anderson and Gonz{\'a}lez, Jorge Wilson and L{\'o}pez, Gabriel Jaime and Isaac, Idi Amin},
  journal={Transactions on Energy Systems and Engineering Applications},
  volume={3},
  number={2},
  pages={1--12},
  year={2022}
}

@article{hu2018design,
  title={Design of visualization plots of industrial alarm and event data for enhanced alarm management},
  author={Hu, Wenkai and Al-Dabbagh, Ahmad W and Chen, Tongwen and Shah, Sirish L},
  journal={Control Engineering Practice},
  volume={79},
  pages={50--64},
  year={2018},
  publisher={Elsevier}
}

@inproceedings{domova2020guiding,
  title={Guiding the Operator's Attention Among a Plurality of Operator Workstation Screens},
  author={Domova, Veronika},
  booktitle={IEEE ETFA'20},
  volume={1},
  pages={541--548},
  year={2020},
  organization={IEEE}
}

@article{noah2014evaluating,
  title={Evaluating alternate visualization techniques for overview displays in process control},
  author={Noah, Benjamin and Kim, Jung-Hyup and Rothrock, Ling and Tharanathan, Anand},
  journal={IIE Transactions on Occupational Ergonomics and Human Factors},
  volume={2},
  number={3-4},
  pages={152--168},
  year={2014},
  publisher={Taylor \& Francis}
}

@article{yang2014evaluation,
  title={Evaluation of visualization techniques for use by facility operators during monitoring tasks},
  author={Yang, Xue and Ergan, Semiha},
  journal={Automation in Construction},
  volume={44},
  pages={103--118},
  year={2014},
  publisher={Elsevier}
}

@inproceedings{fagerlonn2017designing,
  title={Designing a multimodal warning display for an industrial control room},
  author={Fagerl{\"o}nn, Johan and Hammarberg, Kristin and Lindberg, Stefan and Sirkka, Anna and Larsson, Sofia},
  booktitle={Proceedings of the 12th International Audio Mostly Conference},
  pages={1--5},
  year={2017}
}

@article{willett_embedded_2017,
	title = {Embedded {Data} {Representations}},
	volume = {23},
	issn = {1077-2626},
	url = {http://ieeexplore.ieee.org/document/7539328/},
	doi = {10.1109/TVCG.2016.2598608},
	language = {en},
	number = {1},
	urldate = {2022-09-22},
	journal = {IEEE TVCG},
	author = {Willett, Wesley and Jansen, Yvonne and Dragicevic, Pierre},
	month = jan,
	year = {2017},
	pages = {461--470}
}

@article{de2015information,
  title={Information technology in critical care: review of monitoring and data acquisition systems for patient care and research},
  author={De Georgia, Michael A and Kaffashi, Farhad and Jacono, Frank J and Loparo, Kenneth A},
  journal={The Scientific World Journal},
  volume={2015},
  number={1},
  pages={727694},
  year={2015},
  publisher={Wiley Online Library}
}

@inproceedings{mahmoud2003real,
  title={Real-time data acquisition system for monitoring patients in an intensive care unit (ICU)},
  author={Mahmoud, Mohamed},
  booktitle={Multisensor, Multisource Information Fusion: Architectures, Algorithms, and Applications 2003},
  volume={5099},
  pages={320--326},
  year={2003},
  organization={SPIE}
}

@inproceedings{fuchs_evaluation_2013,
	address = {New York, NY, USA},
	title = {Evaluation of alternative glyph designs for time series data in a small multiple setting},
	isbn = {978-1-4503-1899-0},
	url = {https://doi.org/10.1145/2470654.2466443},
	doi = {10.1145/2470654.2466443},
	urldate = {2023-02-09},
    booktitle = {Proceedings of CHI'13},
	publisher = {ACM},
	author = {Fuchs, Johannes and Fischer, Fabian and Mansmann, Florian and Bertini, Enrico and Isenberg, Petra},
	month = apr,
	year = {2013},
	pages = {3237--3246}
}

@inproceedings{mairena_baseline_2020,
 author = {Mairena, Aristides and Dechant, Martin and Gutwin, Carl and Cockburn, Andy},
 title = {A Baseline Study of Emphasis Effects in Information Visualization},
 booktitle = {Proceedings of Graphics Interface 2020},
 series = {GI 2020},
 year = {2020},
 isbn = {978-0-9947868-5-2},
 location = {University of Toronto},
 pages = {327 -- 339},
 numpages = {13},
 doi = {10.20380/GI2020.33},
}

@article{hall_formalizing_2016,
	title = {Formalizing {Emphasis} in {Information} {Visualization}},
	volume = {35},
	issn = {1467-8659},
	url = {https://onlinelibrary.wiley.com/doi/abs/10.1111/cgf.12936},
	doi = {10.1111/cgf.12936},
	language = {en},
	number = {3},
	urldate = {2024-01-06},
	journal = {CGF},
	author = {Hall, K. Wm. and Perin, C. and Kusalik, P. G. and Gutwin, C. and Carpendale, S.},
	year = {2016},
	pages = {717--737}
}

@article{mairena_which_2022,
	title = {Which emphasis technique to use? {Perception} of emphasis techniques with varying distractors, backgrounds, and visualization types},
	volume = {21},
	issn = {1473-8716},
	shorttitle = {Which emphasis technique to use?},
	url = {https://doi.org/10.1177/14738716211045354},
	doi = {10.1177/14738716211045354},
	language = {en},
	number = {2},
	urldate = {2024-01-06},
	journal = {Information Visualization},
	author = {Mairena, Aristides and Gutwin, Carl and Cockburn, Andy},
	month = apr,
	year = {2022},
	note = {Publisher: SAGE Publications},
	pages = {95--129}
}

@article{beecham_faceted_2016,
	title = {Faceted {Views} of {Varying} {Emphasis} ({FaVVEs}): a framework for visualising multi-perspective small multiples},
	volume = {35},
	issn = {1467-8659},
	shorttitle = {Faceted {Views} of {Varying} {Emphasis} ({FaVVEs})},
	url = {https://onlinelibrary.wiley.com/doi/abs/10.1111/cgf.12900},
	doi = {10.1111/cgf.12900},
	language = {en},
	number = {3},
	urldate = {2024-01-06},
	journal = {CGF},
	author = {Beecham, R. and Rooney, C. and Meier, S. and Dykes, J. and Slingsby, A. and Turkay, C. and Wood, J. and Wong, B.l. W.},
	year = {2016},
	pages = {241--249}
}

@inproceedings{vermeulen_informing_2012,
	title = {Informing the design of situated glyphs for a care facility},
	url = {https://ieeexplore.ieee.org/document/6344490},
	doi = {10.1109/VLHCC.2012.6344490},
	urldate = {2024-05-30},
	booktitle = {2012 {IEEE} {Symposium} on {VL}/{HCC}},
	author = {Vermeulen, Jo and Kawsar, Fahim and Simeone, Adalberto L. and Kortuem, Gerd and Luyten, Kris and Coninx, Karin},
	month = sep,
	year = {2012},
	pages = {89--96}
}

@article{hosseinpour_examining_2024,
	title = {Examining {Limits} of {Small} {Multiples}: {Frame} {Quantity} {Impacts} {Judgments} with {Line} {Graphs}},
	issn = {1941-0506},
	shorttitle = {Examining {Limits} of {Small} {Multiples}},
	url = {https://ieeexplore.ieee.org/document/10458324/?arnumber=10458324},
	doi = {10.1109/TVCG.2024.3372620},
	urldate = {2024-07-24},
	journal = {IEEE TVCG},
	author = {Hosseinpour, Helia and Matzen, Laura E. and Divis, Kristin M. and Castro, Spencer C. and Padilla, Lace},
	year = {2024},
	pages = {1--12}
}

@inproceedings{heyer_investigations_2010,
	address = {New York, NY, USA},
	title = {Investigations of {Ubicomp} in the oil and gas industry},
	isbn = {978-1-60558-843-8},
	url = {https://dl.acm.org/doi/10.1145/1864349.1864360},
	doi = {10.1145/1864349.1864360},
	urldate = {2024-08-05},
	booktitle = {Proceedings of {UbiComp}'10},
	publisher = {ACM},
	author = {Heyer, Clint},
	month = sep,
	year = {2010},
	pages = {61--64}
}

@article{brehmer_multi-level_2013,
	title = {A {Multi}-{Level} {Typology} of {Abstract} {Visualization} {Tasks}},
	volume = {19},
	issn = {1941-0506},
	url = {https://ieeexplore.ieee.org/document/6634168},
	doi = {10.1109/TVCG.2013.124},
	number = {12},
	urldate = {2024-08-27},
	journal = {IEEE TVCG},
	author = {Brehmer, Matthew and Munzner, Tamara},
	month = dec,
	year = {2013},
	pages = {2376--2385}
}

@book{munzner_visualization_2015,
	series = {A.{K}. {Peters} visualization series},
	title = {Visualization {Analysis} and {Design}},
	isbn = {978-1-4665-0891-0},
	publisher = {CRC Press, Taylor \& Francis Group},
	author = {Munzner, Tamara},
	year = {2015},
}

@article{javed_graphical_2010,
	title = {Graphical {Perception} of {Multiple} {Time} {Series}},
	volume = {16},
	copyright = {https://ieeexplore.ieee.org/Xplorehelp/downloads/license-information/IEEE.html},
	issn = {1077-2626},
	url = {http://ieeexplore.ieee.org/document/5613429/},
	doi = {10.1109/TVCG.2010.162},
	language = {en},
	number = {6},
	urldate = {2024-09-08},
	journal = {IEEE TVCG},
	author = {Javed, W and McDonnel, B and Elmqvist, N},
	month = nov,
	year = {2010},
	pages = {927--934}
}

@article{healey_attention_2012,
	title = {Attention and {Visual} {Memory} in {Visualization} and {Computer} {Graphics}},
	volume = {18},
	copyright = {https://ieeexplore.ieee.org/Xplorehelp/downloads/license-information/IEEE.html},
	issn = {1077-2626},
	url = {http://ieeexplore.ieee.org/document/5963660/},
	doi = {10.1109/TVCG.2011.127},
	language = {en},
	number = {7},
	urldate = {2024-09-08},
	journal = {IEEE TVCG},
	author = {Healey, C. G. and Enns, J. T.},
	month = jul,
	year = {2012},
	pages = {1170--1188}
}

@article{brehmer_comparative_2020,
author = {M. Brehmer and B. Lee and P. Isenberg and E. Choe},
journal = {IEEE TVCG},
title = {A {Comparative} {Evaluation} of {Animation} and {Small} {Multiples} for {Trend} {Visualization} on {Mobile} {Phones}},
year = {2020},
volume = {26},
number = {01},
issn = {1941-0506},
pages = {364-374},
doi = {10.1109/TVCG.2019.2934397},
publisher = {IEEE Computer Society},
address = {Los Alamitos, CA, USA},
month = {jan}
}

@article{boyandin_qualitative_2012,
	title = {A {Qualitative} {Study} on the {Exploration} of {Temporal} {Changes} in {Flow} {Maps} with {Animation} and {Small}‐{Multiples}},
	volume = {31},
	issn = {0167-7055, 1467-8659},
	url = {https://onlinelibrary.wiley.com/doi/10.1111/j.1467-8659.2012.03093.x},
	doi = {10.1111/j.1467-8659.2012.03093.x},
	language = {en},
	number = {3pt2},
	urldate = {2024-09-09},
	journal = {CGF},
	author = {Boyandin, Ilya and Bertini, Enrico and Lalanne, Denis},
	month = jun,
	year = {2012},
	pages = {1005--1014}
}

@article{robertson_effectiveness_2008,
	title = {Effectiveness of {Animation} in {Trend} {Visualization}},
	volume = {14},
	issn = {1941-0506},
	url = {https://ieeexplore.ieee.org/document/4658146/?arnumber=4658146},
	doi = {10.1109/TVCG.2008.125},
	number = {6},
	urldate = {2024-09-09},
	journal = {IEEE TVCG},
	author = {Robertson, George and Fernandez, Roland and Fisher, Danyel and Lee, Bongshin and Stasko, John},
	month = nov,
	year = {2008},
	pages = {1325--1332}
}

@article{archambault_animation_2011,
	title = {Animation, {Small} {Multiples}, and the {Effect} of {Mental} {Map} {Preservation} in {Dynamic} {Graphs}},
	volume = {17},
	copyright = {https://ieeexplore.ieee.org/Xplorehelp/downloads/license-information/IEEE.html},
	issn = {1077-2626},
	url = {http://ieeexplore.ieee.org/document/5473226/},
	doi = {10.1109/TVCG.2010.78},
	language = {en},
	number = {4},
	urldate = {2024-09-09},
	journal = {IEEE TVCG},
	author = {Archambault, D and Purchase, H and Pinaud, B},
	month = apr,
	year = {2011},
	pages = {539--552}
}

@article{sarikaya_what_2019,
	title = {What {Do} {We} {Talk} {About} {When} {We} {Talk} {About} {Dashboards}?},
	volume = {25},
	issn = {1941-0506},
	url = {https://ieeexplore.ieee.org/document/8443395/?arnumber=8443395},
	doi = {10.1109/TVCG.2018.2864903},
	number = {1},
	urldate = {2024-09-10},
	journal = {IEEE TVCG},
	author = {Sarikaya, Alper and Correll, Michael and Bartram, Lyn and Tory, Melanie and Fisher, Danyel},
	month = jan,
	year = {2019},
	pages = {682--692}
}

@book{wexler_big_2017,
	address = {Hoboken, New Jersey},
	title = {The big book of dashboards: visualizing your data using real-world business scenarios},
	isbn = {978-1-119-28308-9 978-1-119-28271-6},
	shorttitle = {The big book of dashboards},
	language = {eng},
	publisher = {Wiley},
	author = {Wexler, Steve and Shaffer, Jeffrey and Cotgreave, Andy},
	year = {2017}
}

@misc{few_visual_2017,
	title = {Visual {Business} {Intelligence} – {There}’s {Nothing} {Mere} {About} {Semantics}},
	url = {https://www.perceptualedge.com/blog/?p=2793},
	language = {en},
	urldate = {2024-09-10},
	journal = {Perceptual Edge},
	author = {Few, Stephen},
	month = dec,
	year = {2017}
}

@article{edrington_machine_2014,
	title = {Machine {Monitoring} {System} {Based} on {MTConnect} {Technology}},
	volume = {22},
	copyright = {https://www.elsevier.com/tdm/userlicense/1.0/},
	issn = {22128271},
	url = {https://linkinghub.elsevier.com/retrieve/pii/S2212827114009664},
	doi = {10.1016/j.procir.2014.07.148},
	language = {en},
	urldate = {2024-09-10},
	journal = {Procedia CIRP},
	author = {Edrington, Ben and Zhao, Bingyan and Hansel, Adam and Mori, Masahiko and Fujishima, Makoto},
	year = {2014},
	pages = {92--97}
}

@article{andrade_novel_2020,
	title = {Novel {Interface} {Designs} for {Patient} {Monitoring} {Applications} in {Critical} {Care} {Medicine}: {Human} {Factors} {Review}},
	volume = {7},
	issn = {2292-9495},
	shorttitle = {Novel {Interface} {Designs} for {Patient} {Monitoring} {Applications} in {Critical} {Care} {Medicine}},
	url = {https://www.ncbi.nlm.nih.gov/pmc/articles/PMC7367533/},
	doi = {10.2196/15052},
	number = {3},
	urldate = {2024-09-10},
	journal = {JMIR Human Factors},
	author = {Andrade, Evismar and Quinlan, Leo and Harte, Richard and Byrne, Dara and Fallon, Enda and Kelly, Martina and Casey, Siobhan and Kirrane, Frank and O'Connor, Paul and O'Hora, Denis and Scully, Michael and Laffey, John and Pladys, Patrick and Beuch\'{e}e, Alain and \'{O}Laighin, Gear\'{o}id},
	month = jul,
	year = {2020},
	pmid = {32618574},
	pmcid = {PMC7367533},
	pages = {e15052}
}

@article{varpio_helpful_2012,
	title = {The {Helpful} or {Hindering} {Effects} of {In}-{Hospital} {Patient} {Monitor} {Alarms} on {Nurses}: {A} {Qualitative} {Analysis}},
	volume = {30},
	issn = {1538-9774},
	url = {https://journals.lww.com/cinjournal/fulltext/2012/04000/the_helpful_or_hindering_effects_of_in_hospital.6.aspx},
	journal = {CIN: Computers, Informatics, Nursing},
	author = {Varpio, Lara and Kuziemsky, Craig and MacDonald, Charlotte and King, W. James},
	year = {2012},
}

@inproceedings{mastrianni_pop-up_2022,
	address = {New Orleans LA USA},
	title = {”{Pop}-{Up} {Alerts} are the {Bane} of {My} {Existence}”: {Designing} {Alerts} for {Cognitive} {Aids} {Used} in {Time}-{Critical} {Medical} {Settings}},
	isbn = {978-1-4503-9156-6},
	shorttitle = {”{Pop}-{Up} {Alerts} are the {Bane} of {My} {Existence}”},
	url = {https://dl.acm.org/doi/10.1145/3491101.3519669},
	doi = {10.1145/3491101.3519669},
	language = {en},
	urldate = {2023-11-17},
	booktitle = {Proceedings of {CHI'22} {Extended} {Abstracts}},
	publisher = {ACM},
	author = {Mastrianni, Angela and Cui, Hua and Sarcevic, Aleksandra},
	month = apr,
	year = {2022},
	pages = {1--7}
}

@inproceedings{paul_designing_2016,
	address = {San Jose California USA},
	title = {Designing the {Alarm} {Management} {User} {Experience} for {Patient} {Monitoring}},
	isbn = {978-1-4503-4082-3},
	url = {https://dl.acm.org/doi/10.1145/2851581.2851604},
	doi = {10.1145/2851581.2851604},
	language = {en},
	urldate = {2023-11-17},
	booktitle = {Proceedings of CHI'16 Extended Abstracts},
	publisher = {ACM},
	author = {Paul, Sharoda A. and Carroll, Alexander K. and Treacy, Stephen M.},
	month = may,
	year = {2016},
	pages = {710--717}
}

@inproceedings{lee_repeat_2019,
	address = {New York, NY, USA},
	title = {To {Repeat} or {Not} to {Repeat}? {Redesigning} {Repeating} {Auditory} {Alarms} {Based} on {EEG} {Analysis}},
	isbn = {978-1-4503-5970-2},
	shorttitle = {To {Repeat} or {Not} to {Repeat}?},
	url = {https://dl.acm.org/doi/10.1145/3290605.3300743},
	doi = {10.1145/3290605.3300743},
	urldate = {2023-11-17},
	booktitle = {Proceedings of CHI'19},
	publisher = {ACM},
	author = {Lee, Yi-Chen and Cherng, Fu-Yin and King, Jung-Tai and Lin, Wen-Chieh},
	month = may,
	year = {2019},
	pages = {1--10}
}

@article{ancker_effects_2017,
	title = {Effects of workload, work complexity, and repeated alerts on alert fatigue in a clinical decision support system},
	volume = {17},
	issn = {1472-6947},
	url = {https://doi.org/10.1186/s12911-017-0430-8},
	doi = {10.1186/s12911-017-0430-8},
	language = {en},
	number = {1},
	urldate = {2023-11-17},
	journal = {BMC Medical Informatics and Decision Making},
	author = {Ancker, Jessica S. and Edwards, Alison and Nosal, Sarah and Hauser, Diane and Mauer, Elizabeth and Kaushal, Rainu and {with the HITEC Investigators}},
	month = apr,
	year = {2017},
	pages = {36}
}

@Manual{rcoreteam_R_2024,
    title = {R: A Language and Environment for Statistical Computing},
    author = {{R Core Team}},
    organization = {R Foundation for Statistical Computing},
    address = {Vienna, Austria},
    year = {2024},
    url = {https://www.R-project.org/},
}

@Book{wickham_ggplot2_2016,
    author = {Hadley Wickham},
    title = {ggplot2: Elegant Graphics for Data Analysis},
    publisher = {Springer-Verlag NY},
    year = {2016},
    isbn = {978-3-319-24277-4},
    url = {https://ggplot2.tidyverse.org},
}

@book{ware_informationvisualization_2012,
author = {Ware, Colin},
title = {Information Visualization: Perception for Design},
year = {2012},
isbn = {9780123814647},
publisher = {Morgan Kaufmann Publishers Inc.},
address = {San Francisco, CA, USA},
edition = {3}
}

@article{cleveland_graphical_1984,
    title = {Graphical {Perception}: {Theory}, {Experimentation}, and {Application} to the {Development} of {Graphical} {Methods}},
    volume = {79},
    issn = {0162-1459},
    shorttitle = {Graphical {Perception}},
    url = {https://www.jstor.org/stable/2288400},
    doi = {10.2307/2288400},
    number = {387},
    urldate = {2024-02-27},
    journal = {JASA},
    author = {Cleveland, William S. and McGill, Robert},
    year = {1984},
    pages = {531--554},
}

@article{mackinlay_automating_1986,
	title = {Automating the design of graphical presentations of relational information},
	volume = {5},
	issn = {0730-0301},
	url = {https://dl.acm.org/doi/10.1145/22949.22950},
	doi = {10.1145/22949.22950},
	number = {2},
	urldate = {2026-01-15},
	journal = {ACM Trans. Graph.},
	author = {Mackinlay, Jock},
	month = apr,
	year = {1986},
	pages = {110--141},
}

@article{RJ-2017-066,
  author = {Brooks, Mollie E. and Kristensen, Kasper and Benthem, Koen J. van and Magnusson, Arni and Berg, Casper W. and Nielsen, Anders and Skaug, Hans J. and Mächler, Martin and Bolker, Benjamin M.},
  title = {The R Journal: glmmTMB Balances Speed and Flexibility Among Packages for Zero-inflated Generalized Linear Mixed Modeling},
  journal = {The R Journal},
  year = {2017},
  note = {https://doi.org/10.32614/RJ-2017-066},
  doi = {10.32614/RJ-2017-066},
  volume = {9},
  issue = {2},
  issn = {2073-4859},
  pages = {378-400}
}

@article{barry_needs_2026,
	title = {Needs and {Barriers} of {Healthcare} {Teams} to {Implement} {Real}-{Time} {AI} \& {Predictive} {Analytics} in {Cardiothoracic} {ICU}: {A} {Field} {Study}},
	volume = {8},
	issn = {2673-253X},
	shorttitle = {Needs and {Barriers} of {Healthcare} {Teams} to {Implement} {Real}-{Time} {AI} \&amp; {Predictive} {Analytics} in {Cardiothoracic} {ICU}},
	url = {https://www.frontiersin.org/journals/digital-health/articles/10.3389/fdgth.2026.1943663/full},
	doi = {10.3389/fdgth.2026.1943663},
	language = {English},
	urldate = {2026-10-01},
	journal = {Frontiers in Digital Health},
	publisher = {Frontiers},
	author = {Barry, Carey L. and Purdue, Sydney K. and Nourani, Mahsan and Forlines, Clifton and Goodwin, Matthew and Tory, Melanie},
	month = sep,
	year = {2026},
}

\vfill

\end{document}